\documentclass[preprint,12pt]{elsarticle}
\usepackage{epsfig,graphicx,subfigure,epstopdf,color}
\usepackage{amsmath, amssymb,mathabx,mathrsfs}
\usepackage{amsthm}
\usepackage[toc,page]{appendix}
\usepackage{comment}

\newtheorem{thm}{Theorem}[section]

\theoremstyle{definition}

\theoremstyle{remark}
\newtheorem{remark}[thm]{Remark}

\numberwithin{equation}{section}

\theoremstyle{definition}

\newcommand{\eps}{\varepsilon}

\def\N{{\mathbb N}}

\def\R{{\mathbb R}}

\journal{Physica A}

\begin{document}

\begin{frontmatter}

\title{Topological Data Analysis of Financial Time Series: Landscapes of  Crashes}

\author{Marian  Gidea}
\ead{Marian.Gidea@yu.edu}
\address{Yeshiva University, Department of Mathematical Sciences, New York, NY 10016, USA}

\author{Yuri Katz}
\ead{yuri.katz@spglobal.com}
\address{S\&P Global Market Intelligence, 55 Water Str., New York, NY 10040, USA}

\begin{abstract}
We explore the evolution of daily returns of four major US stock market indices during the technology crash of 2000, and the financial crisis of 2007-2009. Our methodology is based on  topological data analysis (TDA). We use  persistence homology to detect and quantify topological patterns that appear in multidimensional time series. Using a sliding window, we extract  time-dependent point cloud data sets, to which we associate a topological space. We detect transient loops that appear in this space, and we measure their persistence. This is encoded in real-valued functions referred to as a 'persistence landscapes'. We quantify the temporal changes in persistence landscapes via their  $L^p$-norms. We test this procedure on multidimensional time series generated by various non-linear and non-equilibrium models. We find that, in the vicinity of financial meltdowns, the  $L^p$-norms exhibit strong growth prior to the primary peak, which ascends during a crash. Remarkably, the average spectral density at low frequencies of the time series of  $L^p$-norms of the persistence landscapes demonstrates a strong rising trend for 250 trading days prior to either dotcom crash on 03/10/2000, or to the Lehman bankruptcy on 09/15/2008. Our study suggests that TDA provides a new type of  econometric analysis, which complements the standard statistical measures. The method can be used to detect early warning signals of imminent market crashes. We believe that this approach can be used beyond the analysis of  financial time series presented here.
\end{abstract}
\end{frontmatter}


\section{Introduction}
Topological Data Analysis (TDA) \cite{Carlsson09,Edelsbrunner10} refers to a combination of statistical, computational, and topological methods allowing to find shape-like structures in data. The TDA has proven  to be a powerful exploratory approach for complex multi-dimensional and noisy datasets. For TDA to be applied, a dataset is encoded as a finite set of points in some metric space. The general and intuitive principle underlying TDA is based on persistence of   $k$-dimensional holes, e.g., connected components $(k =0)$, loops $(k = 1)$, etc., in a topological space that is inferred from random samples for a wide range of scales (resolutions) at which data is looked at. Accordingly, persistent homology is the key topological property under consideration  \cite{Edelsbrunner02,Lum15}.

The procedure to compute persistent homology associated to a point cloud data set involves the construction of a filtration of simplicial complexes, ordered with respect to some resolution (scaling) parameter. As the resolution parameter changes, some topological features appear in the corresponding simplicial complex while others disappear. Thus, each topological feature is assigned a `birth' and a `death' value, and the difference between the two values represents the persistence of that feature. A topological feature that persists for a bigger range of scales can be viewed as a significant one, while a feature that persists for a smaller range can be viewed as a less significant, or a noisy feature. An important quality of the persistence homology method is that it does not require an artificial cutoff between `signal' and `noise'; all topological features that emerge from the data are kept, and assigned `weights' according to their persistence. The output of the filtration procedure is captured in a concise form by a persistence diagram. The two coordinates of each point in the diagram represent the birth value and the death value of a $k$-dimensional hole. An alternative instrument to summarize the information contained in a persistence diagram is a persistence landscape \cite{Bubenik15,Bubenik16}. The latter consists of a sequence of continuous, piecewise linear functions defined in the rescaled birth-death coordinates, which are derived from the persistence diagram. Persistence diagrams have a natural metric space structure whereas persistence landscapes are naturally embedded in a Banach space. Thus, one can study the statistical properties of persistence landscapes, e.g., compute expectations and variances, among other properties  \cite{Bubenik15,Chazal2015}.

A remarkable property of persistence homology is that both persistence diagrams and persistence landscapes are robust under perturbations of the underlying  data. That is, if the data set changes only little, the persistence diagrams/persistence landscapes move only by a small distance. This feature is a key ingredient for mathematically well-founded statistical developments using persistence homology.

Exploration of stable topological structures (or `shapes') in nosy multidimensional datasets has led to new insights, including the discovery of a subgroup of breast cancers \cite{Nicolau11},  is actively used in image processing \cite{Carlsson08}, in signal and time-series analysis \cite{BerwaldG14,BerwaldGVJ15,Perea15a,Munch2016,Marco16,Seversky16}. The latter has primarily been applied to detect and quantify periodic patterns in data \cite{Emrani,Perea15b}, to understand the nature of chaotic attractors in the phase space of complex dynamical systems \cite{Maletic15},  to analyze turbulent flows \cite{Mischaikow-et-al}, and stock correlation networks \cite{Gidea17}.

Motivated by these studies, in this paper, we investigate whether application of TDA to financial time series could help to detect a growing systemic risk in financial markets. Despite the obvious practical interest for policymakers and market participants, prediction of catastrophic market meltdowns is notoriously difficult due to complexity and non-stationarity of the financial system. During last decades, there has been a growing body of empirical and theoretical studies, inspired by analysis of abrupt transitions in complex natural systems, devising  early warning signals (EWS) in financial markets, see, e.g.,
\cite{Guttal16} and references therein. Observations on different markets show that financial crashes are preceded by a period of increasing variance of stock market indices, shifting of spectral density of time series towards low frequencies as well as growing cross-correlations. Yet, there is no consensus regarding the mechanism of financial crises. Moreover, even a relatively short-term forecasting of approaching financial disaster remains one of the open challenges.

We analyze the time-series of daily log-returns of four major US stock market indices: S\&P 500, DJIA, NASDAQ, and Russel 2000. Collectively, these noisy $1D$ signals form a multi-dimensional time series in $4D$-space. We apply a sliding window of certain length $w$ along these time series, thereby obtaining a $4D$-point cloud for each instance of the window. The sliding step is set to one day. Then we compute the  $L^p$-norms ($p = 1$ and $p = 2$) of the persistence landscape of the loops ($1D$ persistent homology) in each of the $4D$-point clouds. The resulting time series of  $L^p$-norms allows to track temporal changes in the state of the equity market. We find that the time series of  $L^p$-norms exhibits strong growth prior to the primary peak, which ascends during a crash. Remarkably, the average spectral density of the time series of the norms of the persistence landscapes demonstrates a strong rising trend at low frequencies for 250 trading days prior to either the dotcom crash on 03/10/2000 or to the Lehman bankruptcy on 09/15/2008. Our study suggests that TDA provides a  new type of
econometric analysis that can be used to detect EWS of imminent market crashes.  The method is very general and can be applied to any asset-types and mixtures of time series.

In Section \ref{sec:background}  we provide a concise and informal review of the TDA background and key concepts employed in this paper.  In Section \ref{sec:turbulence} we compare our method outlined above with the traditional time-delay coordinate embedding procedure. We test the proposed method on several multidimensional time series, generated by various non-linear and non-equilibrium models. These tests help to assess how the characteristics of the underlying process affect the persistence landscapes.  In each case, we let the parameters of the underlying process to change  in specific ways, and observe which features of the signal makes the  $L^p$-norms of the persistence landscapes grow. Section \ref{sec:empirical} presents our findings on financial data, which demonstrate that the time series of the norms of the persistence landscapes and its variability could be used as new EWS of approaching market crashes. Section \ref{sec:conclusions} concludes the paper.

\section{Background}\label{sec:background}
TDA uses topology to extract information from noisy data sets. An important step is to assign topological invariants that remain robust under small changes of the scale at which the data set is being considered. This is particularly relevant for analyzing complex systems, when there is no information available on the underlying stochastic process. The method of persistence homology, which we use in this paper, allows computing the topological features at all scales, and ranking these features according to the range of scales at which these features can be observed.

Here we provide a brief account of persistent homology, with an emphasis on the tools that we use directly. For a rigorous exposition, we refer the interested reader to the textbook \cite{Edelsbrunner10}.

The input data that we analyze is a point cloud data set consisting, of a family of points  $X = \{x_1, \ldots ,x_n\}$ in an Euclidean space $\mathbb{R}^d$. We associate a topological space to such data set as follows. For each distance $\eps>0$ we define the so called
Vietoris-Rips simplicial complex $R(X, \eps)$, or, simply Rips complex, obtained as follows:\begin{itemize}\item for each $k=0,1,2, \ldots$, a $k$-simplex of vertices $\{x_{i_1},\ldots, x_{i_k}\}$ is part of $R(X,\eps)$ if and only if the mutual distance between  any pair of its   vertices is less than~$\eps$, that is \[d(x_{i_j},x_{i_l})<\eps, \textrm { for all } x_{i_j},x_{i_l}\in \{x_{i_1},\ldots, x_{i_k}\}. \]\end{itemize} If we think of $\eps$ as the resolution at which that data is looked at, then a $k$-simplex is included in $R(X,\eps)$   for every set of  $k$ data points that are indistinguishable from one another at   resolution level $\eps$.

The Rips simplicial complexes $R(X, \eps)$ form a filtration, that is, $R(X,\eps)\subseteq R(X,\eps')$ whenever $\eps<\eps'$. For each such complex, we can compute its $k$-dimensional homology $H_k(R(X,\eps)$ with coefficients in some field. Informally, the generators of the $0$-dimensional homology group $H_0(R(X,\eps)$  correspond to the connected components of $R(X,\eps)$, the generators of the $1$-dimensional homology group  $H_1(R(X,\eps))$ correspond to the  `independent loops' in $R(X,\eps)$,  the generators of the $2$-dimensional homology group $H_0(R(X,\eps))$ correspond to  `independent cavities' in $R(X,\eps)$, etc.

In the sequel, we will use only the $1$-dimensional homology. As an  example of a loop at level $\eps$, consider a set of  $4$ points $\{x_{i_0}, x_{i_1}, x_{i_2}, x_{i_3}\}$ in $X$ such that the distance $d(x_{i_j},x_{i_{j+1}})<\eps$, for $j=0,\ldots,3$, with $j$ computed mod $4$, but no other pair of points in this set is within a distance of $\eps$. We note that, by the definition of the Rips complex, we cannot have a loop consisting of only $3$-points.

The filtration property of the Rips complexes induces a filtration on the corresponding homologies, that is
$H_k(R(X,\eps))\subseteq H_k(R(x,\eps'))$ whenever $\eps<\eps'$, for each $k$. These inclusions determine canonical homomorphisms $H_k(R(X,\eps))\hookrightarrow H_k(R(x,\eps'))$, for $\eps<\eps'$. Due to this family of induced mappings,  for each non-zero $k$-dimensional homology class $\alpha$  there exists a pair of  values $\eps_1<\eps_2$, such that $\alpha \in H_k(R(X,\eps_1))$ but $\alpha$ is not in the image of any $H_k(R(X,\eps_1-\delta))$, for $\delta>0$, and the  image of $\alpha$ in $H_k(R(X,\eps'))$ is non-zero for all $\eps_1< \eps'<\eps_2$, but the  image of $\alpha$ in $H_k(R(X,\eps_2))$ is zero. In this case, one says that $\alpha$ is `born' at the parameter value $b_\alpha:=\eps_1$, and `dies' at the parameter value $d_\alpha=\eps_2$, or that the pair $(b_\alpha, d_\alpha)$ represent the `birth' and `death' indices of $\alpha$. The multiplicity
$\mu_\alpha(b_\alpha,d_\alpha)$ of the point $(b_\alpha,d_\alpha)$ equals the number of classes $\alpha$ that are born at ${b_\alpha}$ and die at ${d_\alpha}$.
This multiplicity is finite since the simplicial complex is finite.

\begin{figure}
\centering
\includegraphics[width=0.85\textwidth]{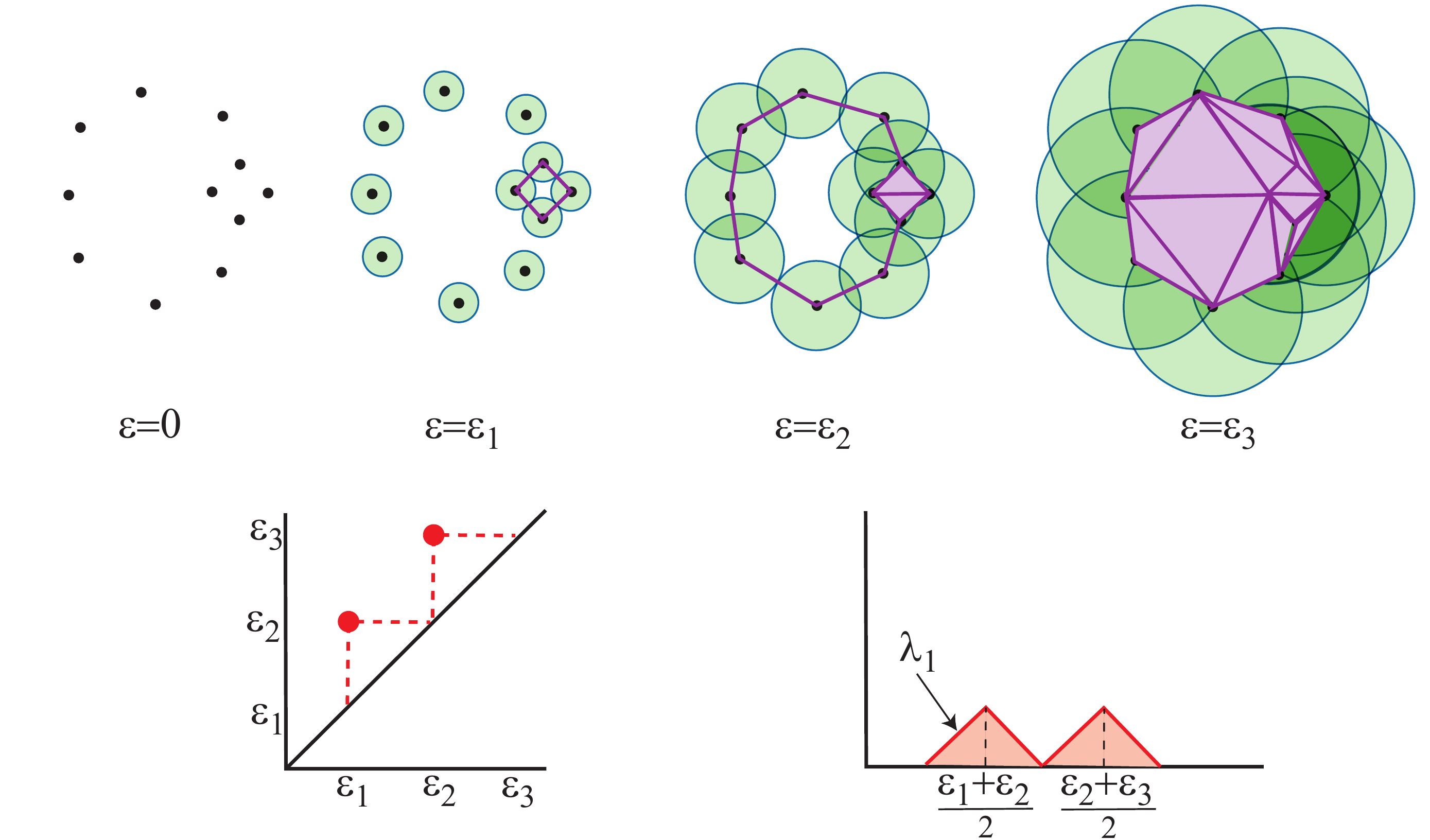}
\caption{Rips filtration of simplicial complexes  illustrating the birth and death of loops; the $1$-dimensional persistence diagram and the corresponding persistence landscape are shown below. }
\label{fig:circles_persistence.png}
\end{figure}

The information on  the $k$-dimensional homology generators at all scales can be encoded in a so called persistence diagram $P_k$. Such a diagram consists of:
\begin{itemize}
\item  For each $k$-dimensional homology class $\alpha$ one assigns a point $z_\alpha=(b_\alpha,d_\alpha)\in\mathbb{R}^2$ together with its multiplicity $\mu_\alpha(b_\alpha,d_\alpha)$;
%\textcolor{red}{this is incorrect needs to be fixed}
%\footnote{Note that since $w_j^*\geq w_l^*$ we adopt the non-standard convention to designate the first coordinate corresponding to death and the second coordinate corresponding to birth.}
\item In addition, $P_k$ contains all points in the positive diagonal  of $\mathbb{R}^2$. These points represent  all trivial homology generators that are born and instantly die at every level; each point on the diagonal has infinite multiplicity.
\end{itemize}
The axes of a persistence diagram are birth indices on the horizontal axis and death indices on the vertical axis.
For an example, see Fig. \ref{fig:circles_persistence.png}.

The space of persistence diagrams can be endowed with a metric space structure.
A standard metric that can be used is  the \emph{degree $p$ Wasserstein distance}, with $p\geq 1$, defined by
\[ W_{p}( P^1_k, P^2_k)=  \inf_{\phi}\left[\sum_{q\in P^1_k} \|x- \phi(x)\|^p_\infty\right]^{1/p},\]
where the summation is  over all bijections $\phi: P^1_k\to P^2_k$, and $\|\cdot\|_\infty$ denotes the $\sup$ norm. Since persistence diagrams include the diagonal set, the above summation includes pairing of points $(x,\phi(x))$ between off-diagonal points and diagonal points in $P^1_k$, $P^2_k$. When $p=\infty$ the Wasserstein distance $W_\infty$  is known as the `bottleneck' distance.

A remarkable property that makes persistence homology suitable to analyze noisy data is its \emph{robustness} under small perturbations. Informally, this property says that if the underlying cloud point data set changes only `little', then the  corresponding  persistence diagrams moves only a `small' Wasserstein distance from the persistence diagram
corresponding to the unperturbed data; see \cite{Cohen-Steiner}.

The space of persistence diagrams $\mathscr{P}$ endowed with the Wasserstein distance forms a metric space $(\mathscr{P}, W_p)$. However, this metric space is not complete, hence not appropriate for statistical treatment.
This is an inconvenience for our purposes, since we want to study time series of persistence diagrams with statistical tools.

To address this issue, several works have been devoted to either modify the structure of the space of persistence diagrams endowed so that it has a more amenable structure (e.g., of a Polish space, or geodesic space), or to embed the space of persistence diagrams into function spaces. One such an embedding is based on persistence landscapes, consisting of sequences of functions in the  Banach space $L^p(\N\times \R)$; see  \cite{Bubenik15,Bubenik16}.
We now define persistence landscapes. For each birth-death point $(b_\alpha,d_\alpha) \in P_k$, we first define a piecewise linear function
\begin{equation}\label{eqn:landscape_1}
f_{(b_\alpha,d_\alpha)}=\left\{
                          \begin{array}{ll}
                            x-b_\alpha, & \hbox{if $x\in\left(b_\alpha,\frac{b_\alpha+d_\alpha}{2}\right]$;} \\
                            -x+d_\alpha, & \hbox{if $x\in\left(\frac{b_\alpha+d_\alpha}{2},d_\alpha\right)$;} \\
                            0, & \hbox{if $x\not\in (b_\alpha, d_\alpha)$.}
                          \end{array}
                        \right.
\end{equation}

To avoid confusion, we point out that in  Eq. \eqref{eqn:landscape_1}, by $(b_\alpha,d_\alpha)$ we mean the open interval with endpoints $b_\alpha$ and $d_\alpha$, while earlier we also denoted by $(b_\alpha,d_\alpha)$ a point in $P_k$ of coordinates $b_\alpha$ and $d_\alpha$.

To a persistence diagram $P_k$ consisting of a finite number of off-diagonal points, we associate
a sequence of functions $\lambda=(\lambda_k)_{k\in\mathbb{N}}$, where $\lambda_k:\mathbb{R}\to[0;1]$ is given by
\begin{equation}\label{eqn:landscape_2} \lambda_k(x)=k\textrm{-max}\{f_{(b_\alpha,d_\alpha)}(x)\,|\, (b_\alpha,d_\alpha)\in P_k\}
\end{equation}
where the $k\textrm{-max}$ denotes the $k$-th largest value of a  function.
We set $\lambda_k(x) = 0$ if
the $k$-th largest value does not exist. For an example of a persistence landscape, see Fig.~\ref{fig:circles_persistence.png}.

Thus, persistence landscapes form a subset of the Banach space $L^p(\N\times \R)$. This consists of sequences of functions $\eta =(\eta_k)_{k\geq 0}$, where $\eta_k:\R\to\R$ for $k\geq 0$. This set  has an obvious vector space structure, given by $(\eta^1+\eta^2)_k(x)=\eta^1_k(x)+\eta^2_k(x)$, and $(c\cdot\eta)_k(x)=c\cdot\eta_k(x)$, for all $x\in\mathbb{R}$, and $k=1,2,\ldots$ It  becomes a Banach  space when endowed with the norm
\begin{equation}\label{eqn:landscape_3}\|\eta\|_p=\left (\sum_{k=1}^{\infty}\|\eta_k\|_p^p\right)^{1/p},\end{equation}
where $\|\cdot\|_p$ denotes the $L^p$-norm, i.e., $\|f\|_p=\left ( \int _\mathbb{R} |f|^p\right )^{1/p}$, where the integration is with respect with  the Lebesgue measure on $\R$.
In the sequel, we will use only the $L^1$ and $L^2$ norms.
Generally, it is not possible to go back and forth between persistence diagrams and landscapes. There exist landscapes that do not correspond to persistence diagrams, e.g., if $\lambda^1$, $\lambda^2$ are two persistence landscapes derived, via Eq. \eqref{eqn:landscape_1}, from corresponding  persistence diagrams $P^1,P^2\in\mathscr{P}$, respectively,  the mean $\bar\lambda=(\lambda^1+\lambda^2)/2$ may not correspond, via  Eq. \eqref{eqn:landscape_1}, to any persistence diagram $\bar P\in\mathscr{P}$; see \cite{Bubenik15}.
However, the Banach space structure makes persistence landscapes suitable for treatment via \emph{statistical
methods}.

\section{Description of the  method and testing on synthetic time series}\label{sec:turbulence}
In Section \ref{sec:empirical} we will analyze financial time series using persistence landscapes and
their   norms ($p = 1$ and $p = 2$). We will see that these norms exhibit substantial growth prior to financial crashes, while they exhibit a tame behavior when the market is stable.
In this section, we will explain our method of analysis and test it on several synthetic multidimensional time series generated by various non-linear and non-equilibrium models. The objective is to assess how various features of the signal affect the growth of the $L^p$ -norms of the persistence landscapes.

\subsection{Method}
Consider $d$ time series  $\{x^k_n\}_n$, where $k=1,\ldots, d$, and a sliding window of size $w$.  For each instance $t_n$, we have a point $x(t_n)=(x^1_n,\ldots, x^d_n)\in \mathbb{R}^d$. Then for each time-window of size $w$ we have a point cloud data set consisting of $w$ points in $\mathbb{R}^d$, namely  $X_n=(x(t_n), x(t_{n+1}), \ldots, x(t_{n+w-1}))$. Algorithmically, the point cloud is formed by a $d\times w$  matrix, where $d$ is the number of columns corresponding to the number of $1D$ time series under study and $w$ is the size of the window, which determines the length of a column. Then TDA is applied on top of the time-ordered sequence of point clouds to study the time-varying topological properties of the multidimensional time series, from \emph{window} to \emph{window}. For each point cloud we compute the persistence diagram of the Rips filtration, the corresponding persistence landscape, and its  $L^p$-norms. We use efficient computational algorithms and the relevant TDA libraries allowing computation of topological features from noisy observations \cite{Morozov14}. In our implementation we employ the R-package ``TDA" \cite{Fasy16}, which greatly simplifies the application programming interface.

In the following subsection we illustrate our method to analyze a time series generated by a  stochastic version of the  H\'enon map.

\begin{remark} We now compare our approach with the well-known method based on time-delay coordinate embedding \cite{Takens}, which is broadly used in the conventional analysis and TDA studies of $1D$ signals.

First, we describe the  case when the underlying signal is stationary and represents the time series of a chaotic attractor (hidden in a `black box'). A sliding window of size $m$ is applied to this time series, with each instance of a window giving rise to a point in $\mathbb{R}^m$. The theory ensures the resulting set of points in $\mathbb{R}^m$ represents  (approximately) an embedding of the attractor in the $m$-dimensional space, provided the dimension $m$ is chosen large enough: at least twice the fractal dimension of the attractor \cite{Sauer}.  There are several methods to estimate, directly from the time series, the minimal dimension $m$ necessary to achieve the embedding \cite{Abarbanel,Packard}. We should stress that this approach becomes increasingly difficult to apply when the dimension of the attractor is large, or in the presence of noise \cite{Muldoon}; see also   \cite{Mischaikow97}.

The second case that we describe is when the $1D$ time series is associated to a non-stationary process; we follow \cite{Seversky16}. Given a lengthy $1D$ time-series,  one  takes a partition of  it into multiple segments of a predefined size $w$. Within each segment, a sliding window of size $m$ is applied to the time series, where each instance of the window is consists of a sequence of $m$ consecutive data-points from the time series, where $m\ll w$.
Again, $m$ has to be chosen suitably large. The time-ordered sequence of sliding windows for each segment forms a single point cloud of size $(w-m+1)$ in $\mathbb{R}^m$. Thereby, one obtains a representation of each segment of length $w$ of the $1D$ dataset in the $m$-dimensional space. Then one applies TDA   on top of the embedded point clouds to study the time-varying topological properties of the time series, \emph{from segment to segment}.

We stress that our method is quite different from the time-delay embedding method. As we do not aim to obtain an embedding of the time series in some \emph{high dimensional space} -- as such embedding might be practically impossible to realize, due to the a priori lack of an attractor, and to the intrinsic stochasticity of the time series -- we choose to represent the time series in a low dimensional space, namely in $\mathbb{R}^d$, where $d \ll w$, as we are dealing with a handful (that is, $d$) separate time series. Our approach has only one parameter, the size $w$ of the window sliding along the time series, as in contrast with the time-delay method, which has two parameters, the size $w$ of the segments, and the size $m$ of a sliding window along each segment.

We stress that both the length of the segment  $w$ in \cite{Seversky16} and the length of the sliding window  $w$ in our method play the same role, namely to determine the size of the point cloud to which we apply TDA, from one segment to another in the former, and from one window to another in the latter. The dimension $m$ of the point cloud  is derived from the time-delay embedding procedure in the former, and is naturally given by the number $d$ of time series under consideration in the latter.
\end{remark}

\subsection{Chaotic time series with noise}
First, we recall the   H\'enon map   \cite{Henon}, which  can be viewed as a $2$-dimensional analogue of the well known logistic map. The latter has been used as a paradigm for economic cycles (see, e.g., \cite{Ausloos}). The evolution of stable states, bifurcations, periodic and chaotic regimes -- displayed by both the logistic map and the H\'enon map --  provides a useful analogy for the dynamics of economic oscillations and financial crashes (see, e.g., \cite{Sornette}).
A model for financial crashes based on bifurcations in dynamical systems appears in \cite{Douady}.

The equations of the H\'enon map are:
\begin{equation}\label{eqn:henon}\begin{split}x_{n+1} &= 1-a x_n^2+b y_n,\\
y_{n+1} &= x_{n+1}.
\end{split}\end{equation}
where $a$, $b$ ar real parameters. It has been shown that, for some range of values of $a$, $b$, every initial condition $(x_0,y_0)$ in some appropriate region of the plane -- the `basin of attraction' --, the sequence $(x_n,y_n)$ approaches the same invariant set, known as the H\'enon attractor. For some values of $a$ and $b$, the attractor is \emph{chaotic}.
Let us fix the parameter $b=0.3$ and let $0\leq a\leq 1.4$. Numerical experiments show that for fixed values of parameter the parameter $a$ with $0<a<1.06$, there exists an  attractor that undergoes period-doubling bifurcations, whereas at $a\approx 1.06$ a  chaotic attractor  emerges; for  $a$ between $1.06$ and $1.4$, there exist intervals of values of $a$ for  which  there is a chaotic attractor,  which are interspersed with intervals of values of $a$ for which there is an attractive periodic orbit.
%Bifurcation diagram to be included?

\begin{figure}
\centering
\includegraphics[width=0.85\textwidth]{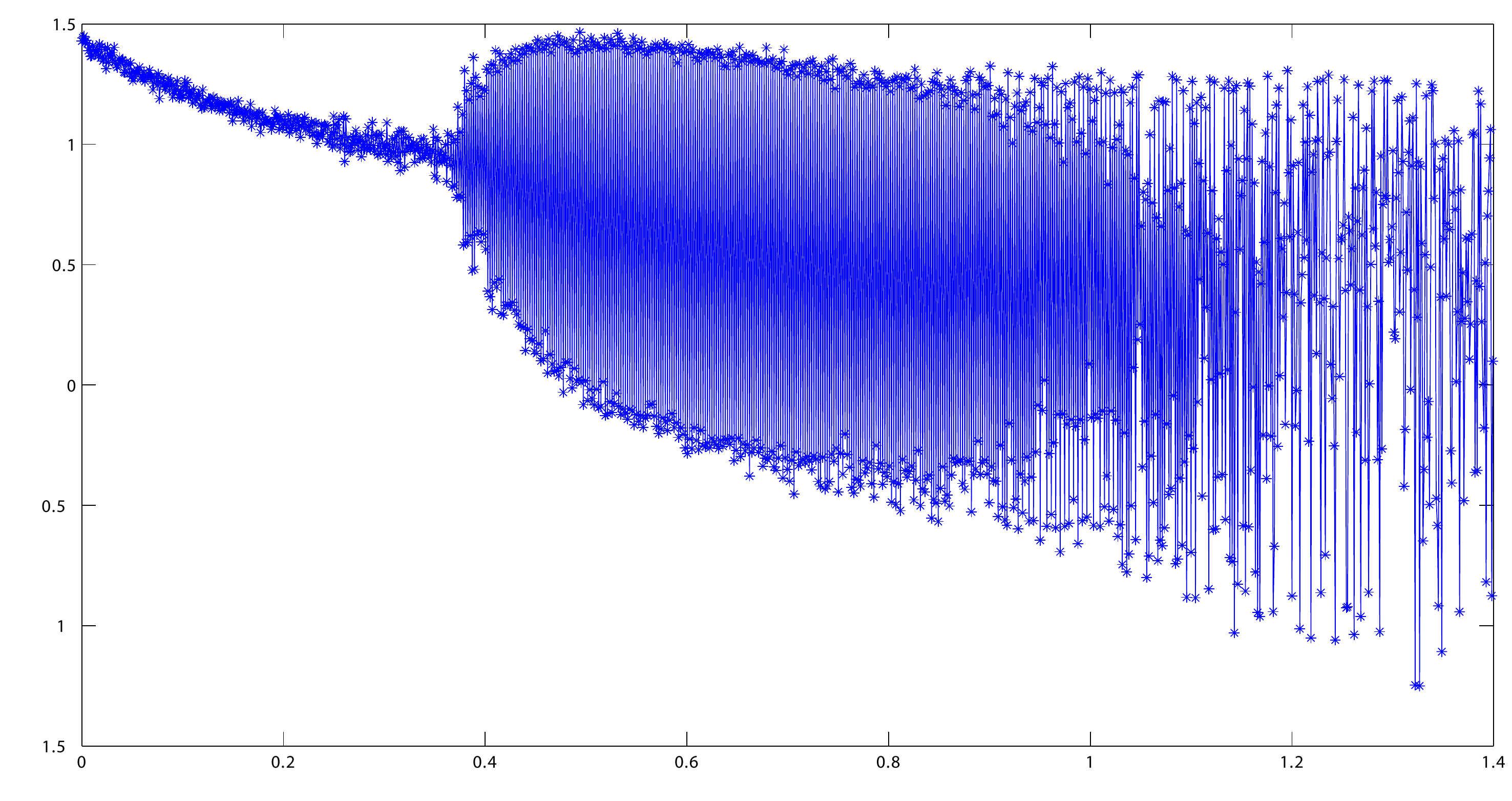}
\caption{An $x$-time series corresponding to the system  Eq. \eqref{eqn:noisy_henon}. The horizontal axis corresponds to the values of the parameter $a$.}
\label{fig:gen_henon_2_d_1.pdf}
\end{figure}

We modify the equations of the H\'enon attractor by making one of the  parameters change slowly in time (analogous to an external forcing), and we also  dress it with noise.
For  fixed values of the parameter $b$ ($b=0.27$, $b=0.28$, $b=0.29$, and $b=0.3$), we let the parameter $a$ grow slowly and linearly in time, from $a=0$  to $a=1.4$, and also  add the Gaussian noise. That is, we consider the system:
\begin{equation}\label{eqn:noisy_henon}\begin{split}x_{n+1} &= 1-a_n x_n^2+b y_n+\sigma W_n \sqrt{\Delta t} ,\\
y_{n+1} &= x_{n+1} +\sigma W_n \sqrt{\Delta t},\\
a_{n+1}&=a_n+\Delta t,
\end{split}\end{equation}
where $W_n$ is a normal random variable,  $\Delta t>0$ is a small step size, and $\sigma>0$ is the noise intensity.
The stochastic terms in  Eq. \eqref{eqn:noisy_henon} correspond to a diffusion process. The time changes according to $t_{n+1}=t_n+\Delta t$, hence $a_n$ can be viewed as equivalent to the time variable.
An realization of the $x$-time series for this system is shown in Fig.~\ref{fig:gen_henon_2_d_1.pdf}.

\begin{figure}
\centering$\begin{array}{cc}
\includegraphics[height=0.224\textwidth]{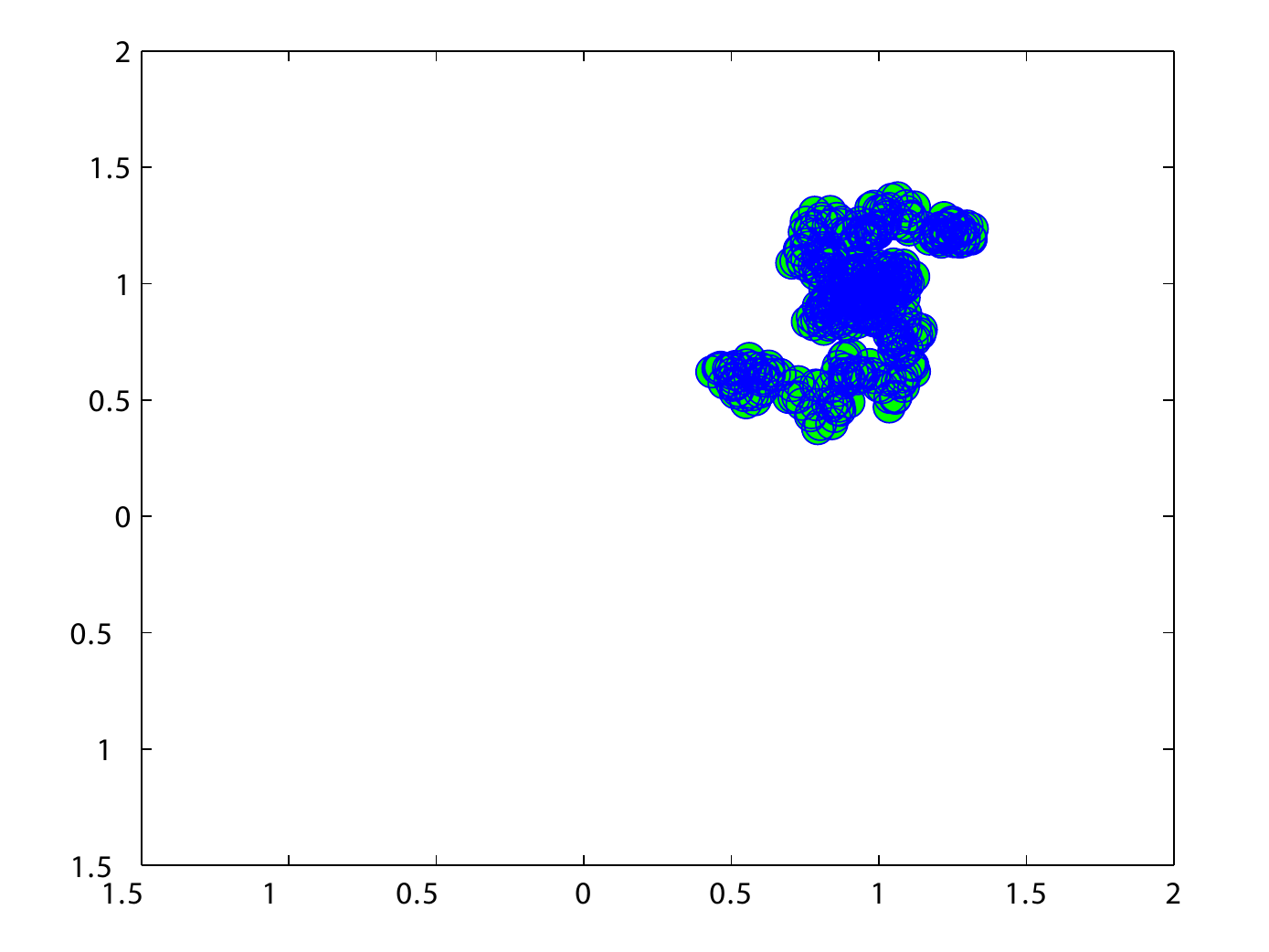}& \includegraphics[height=0.224\textwidth]{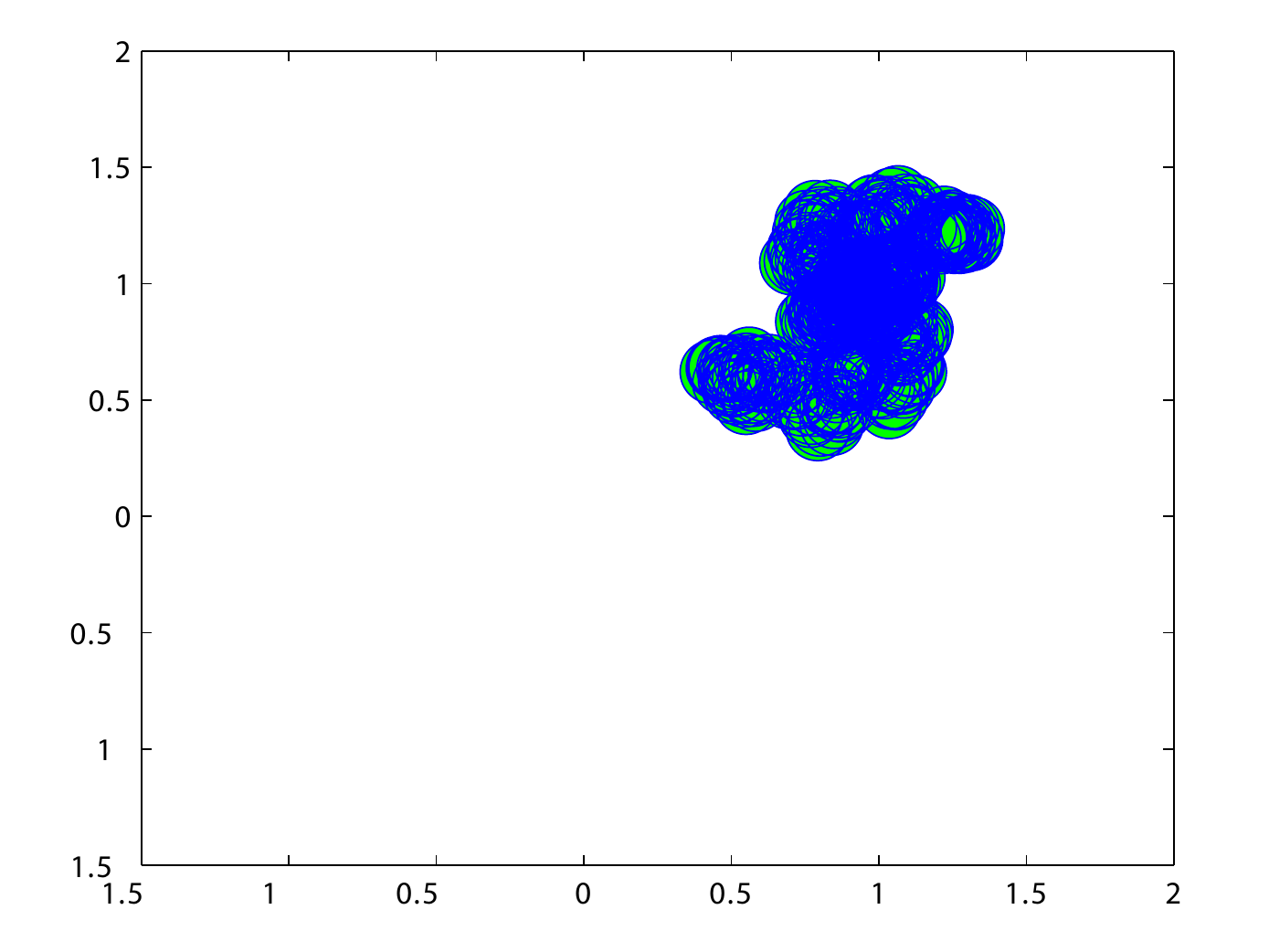}\\
\includegraphics[height=0.224\textwidth]{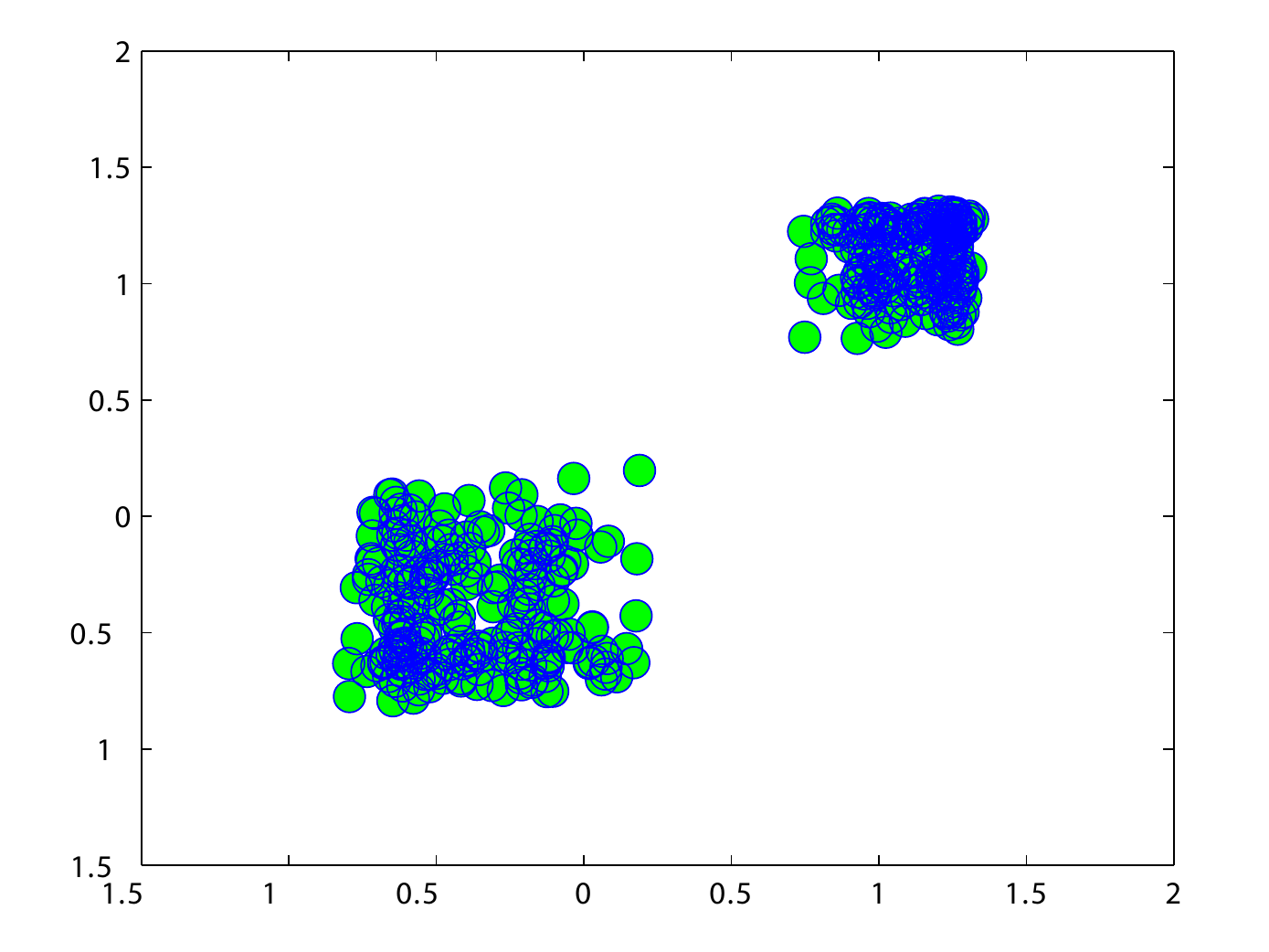}& \includegraphics[height=0.224\textwidth]{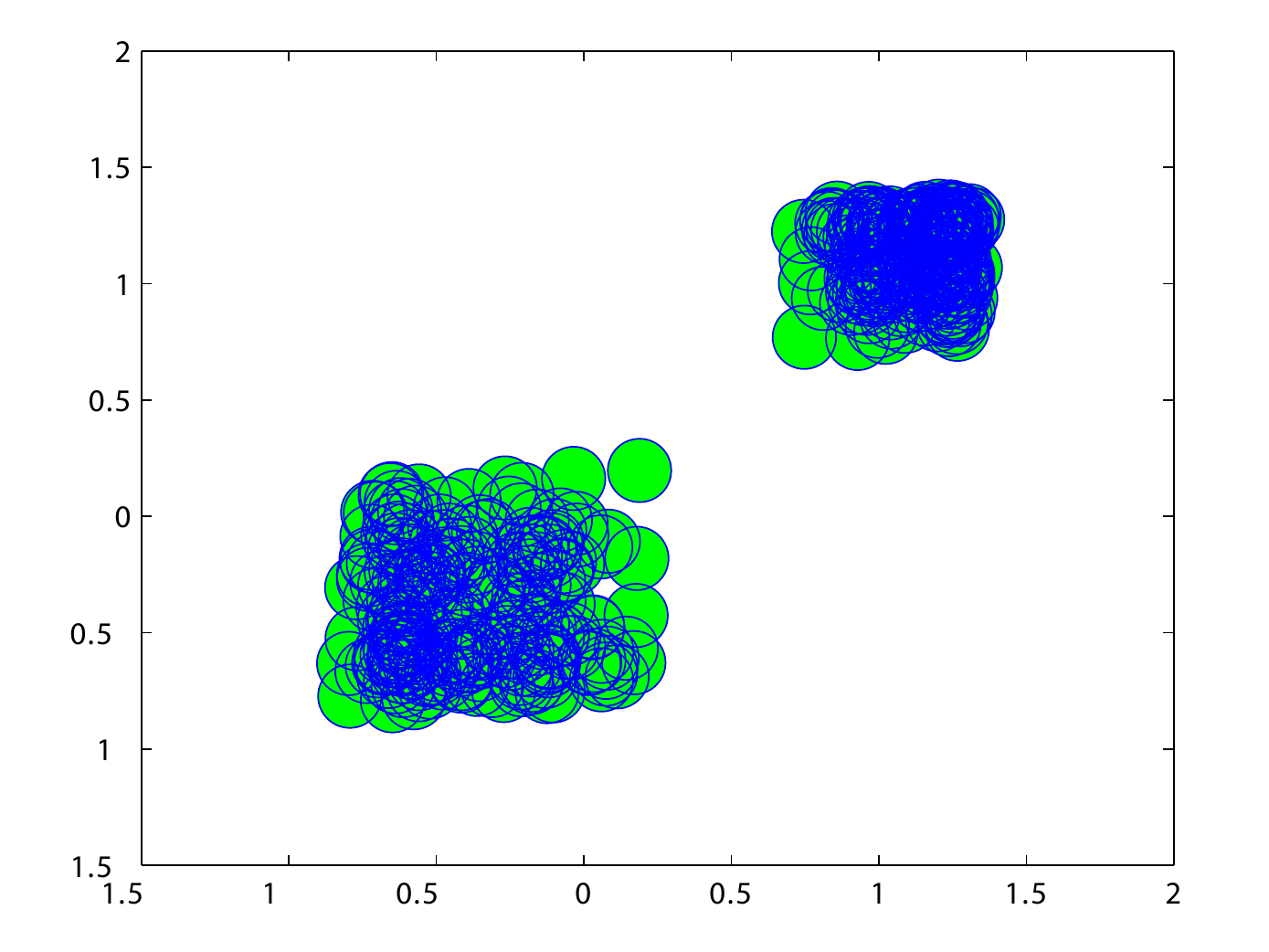}\\
\includegraphics[height=0.224\textwidth]{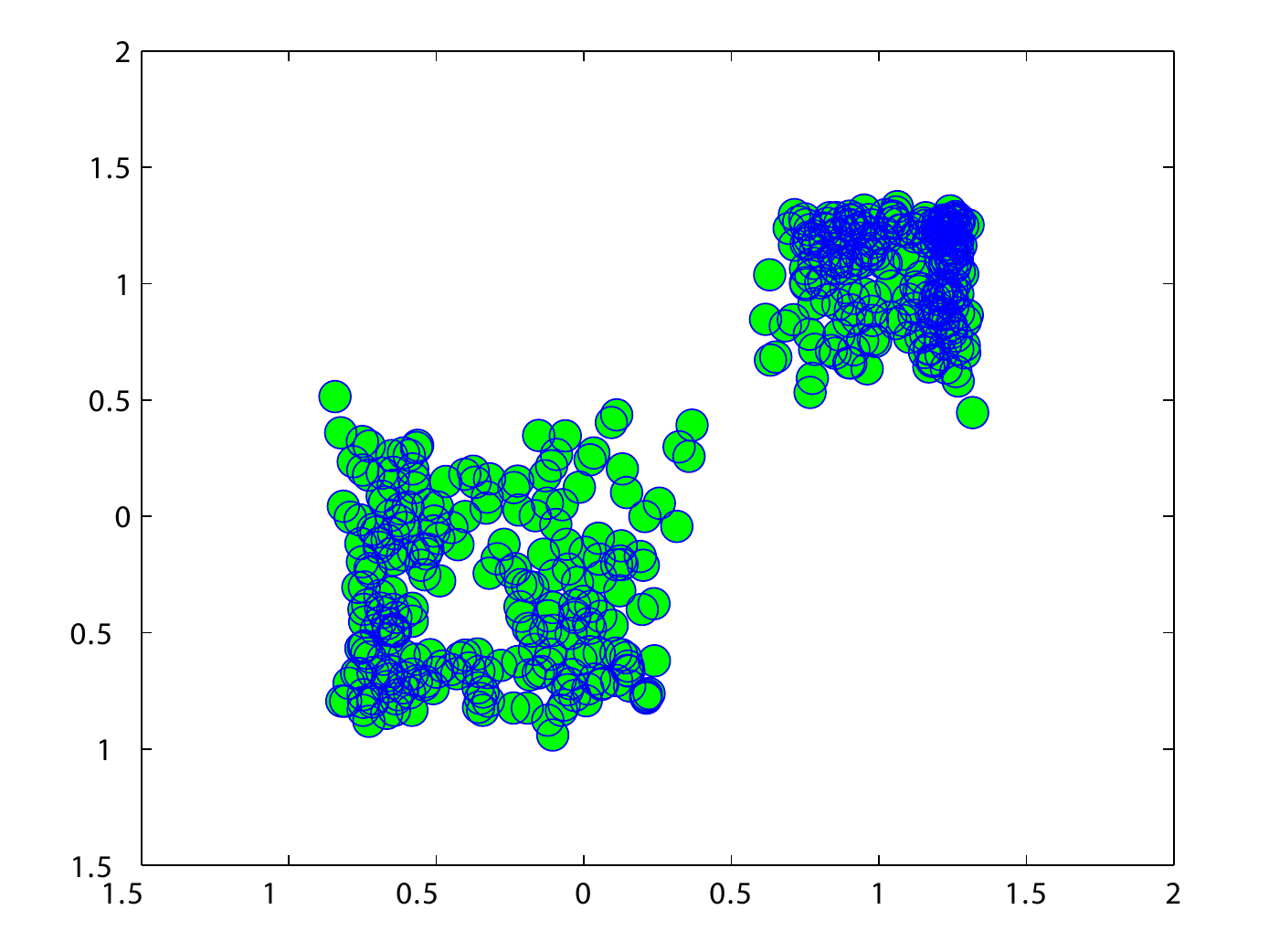}& \includegraphics[height=0.224\textwidth]{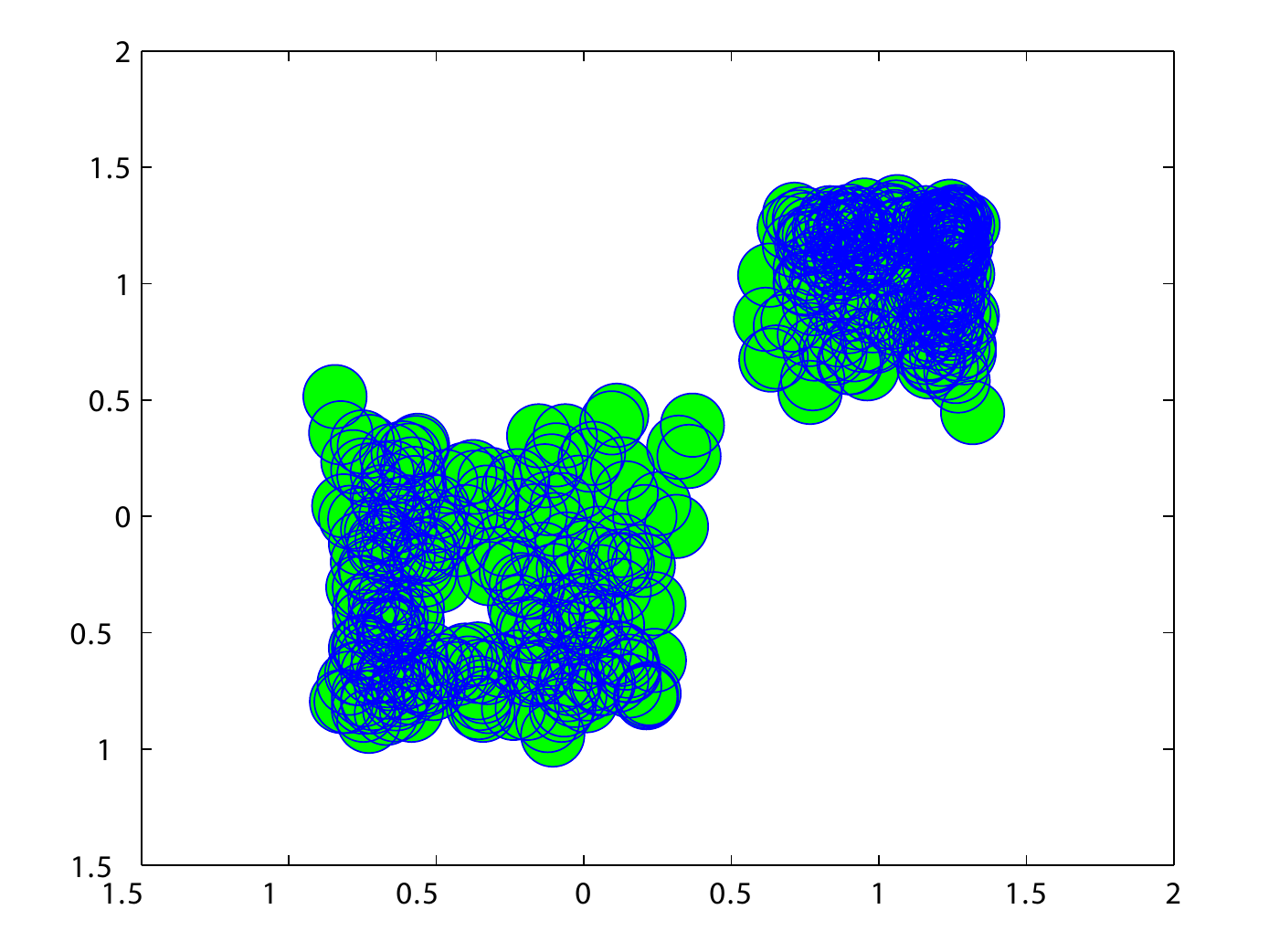}\\
\includegraphics[height=0.224\textwidth]{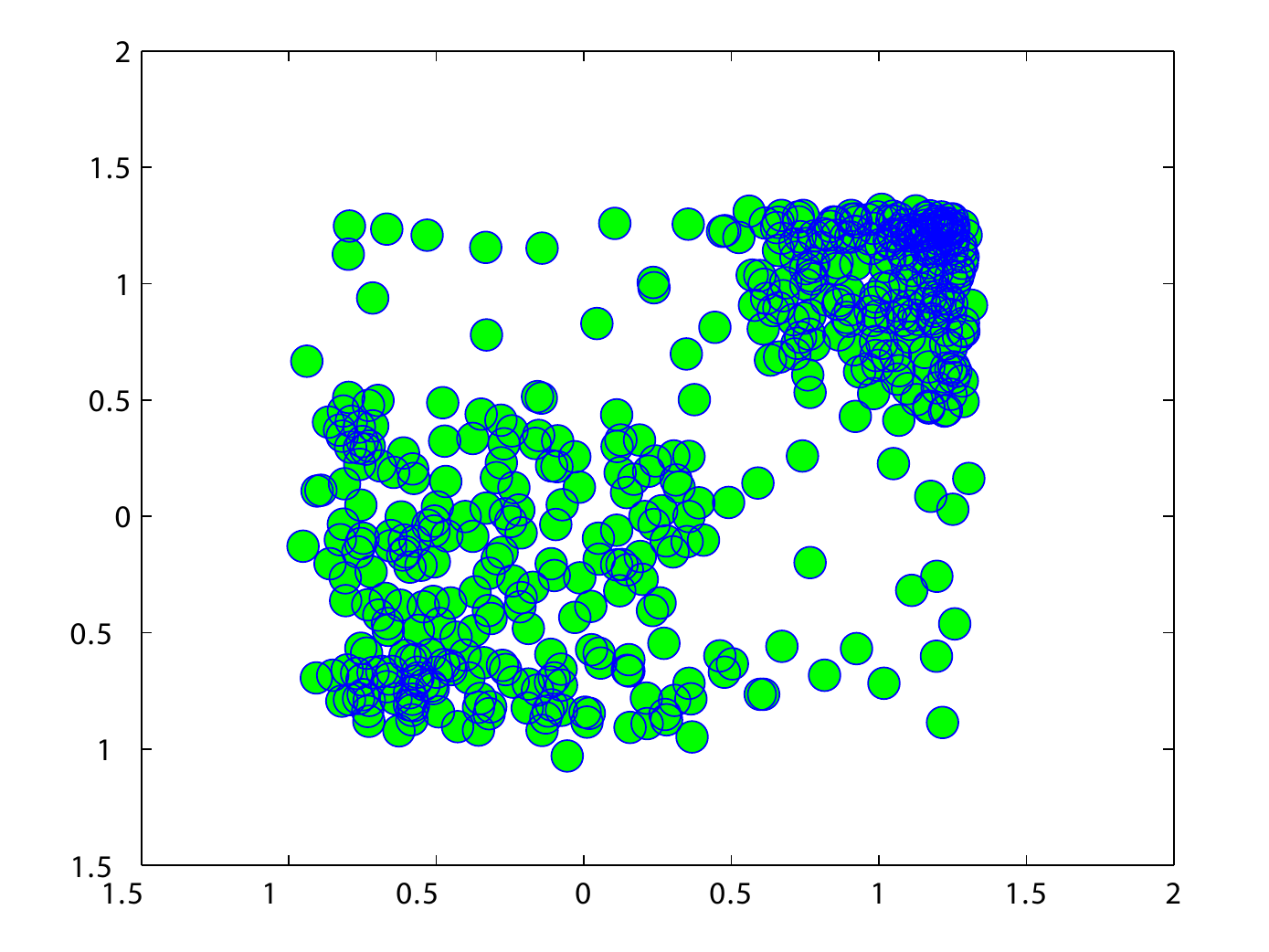}& \includegraphics[height=0.224\textwidth]{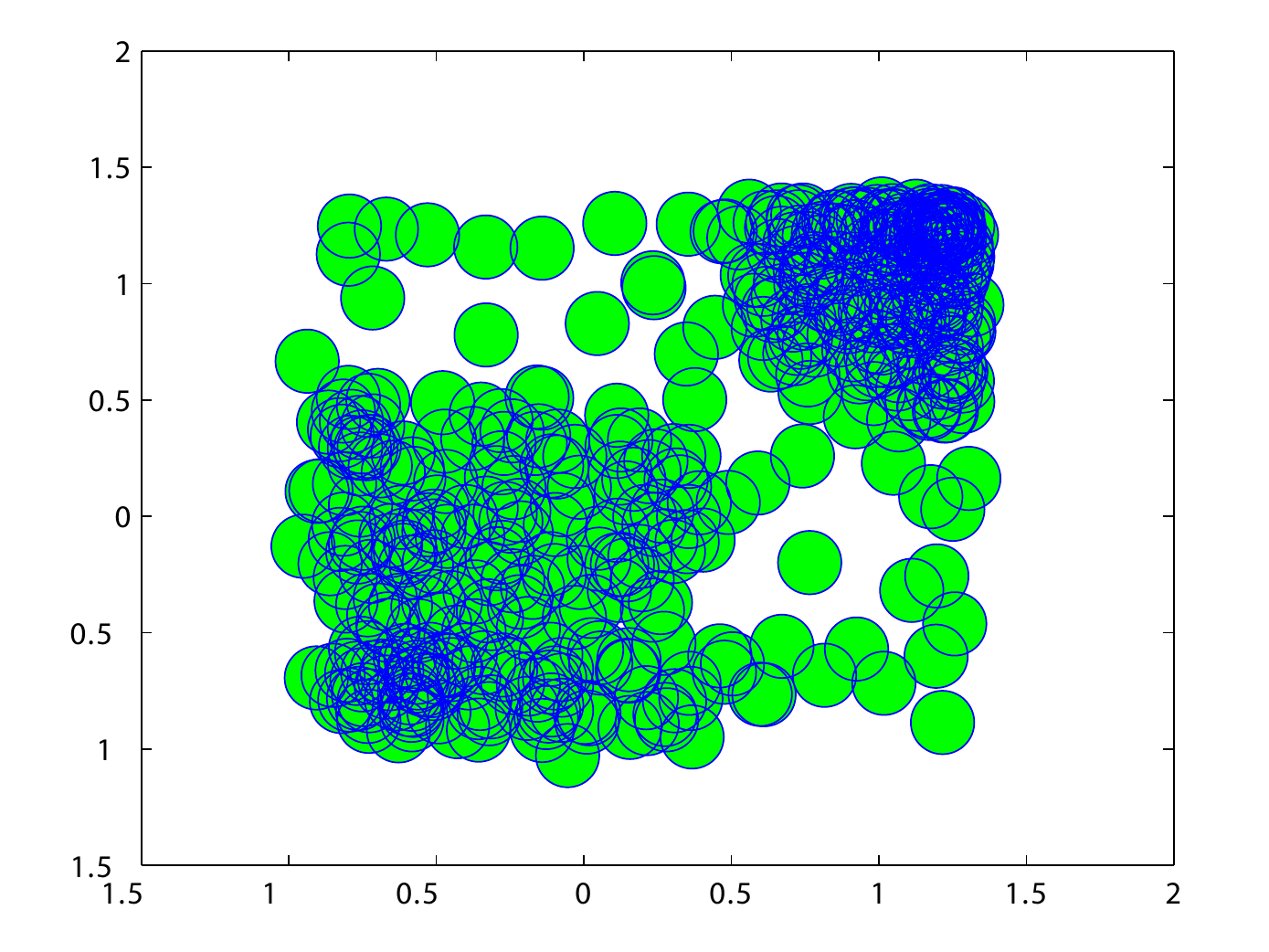}\\
\includegraphics[height=0.224\textwidth]{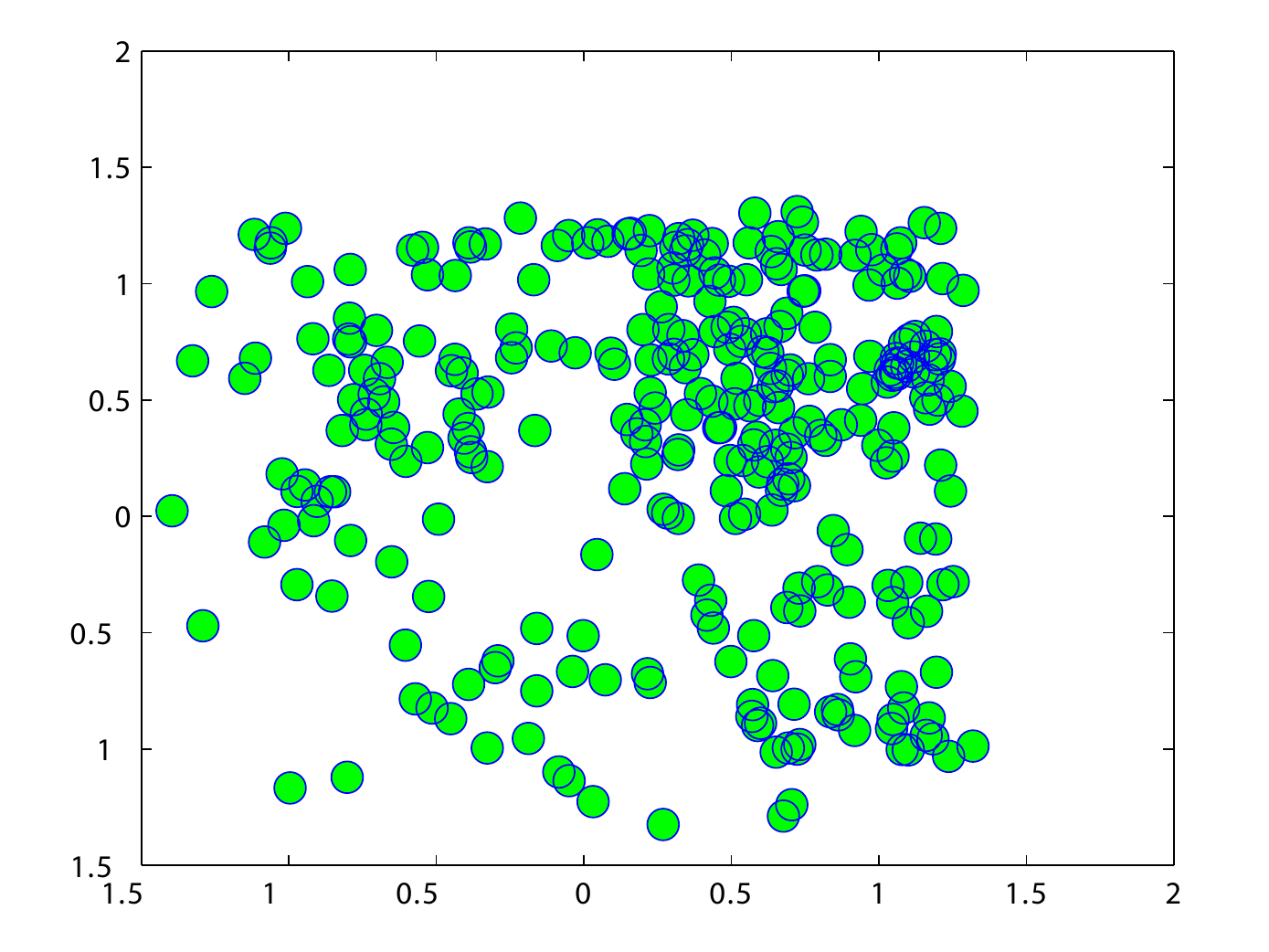}& \includegraphics[height=0.224\textwidth]{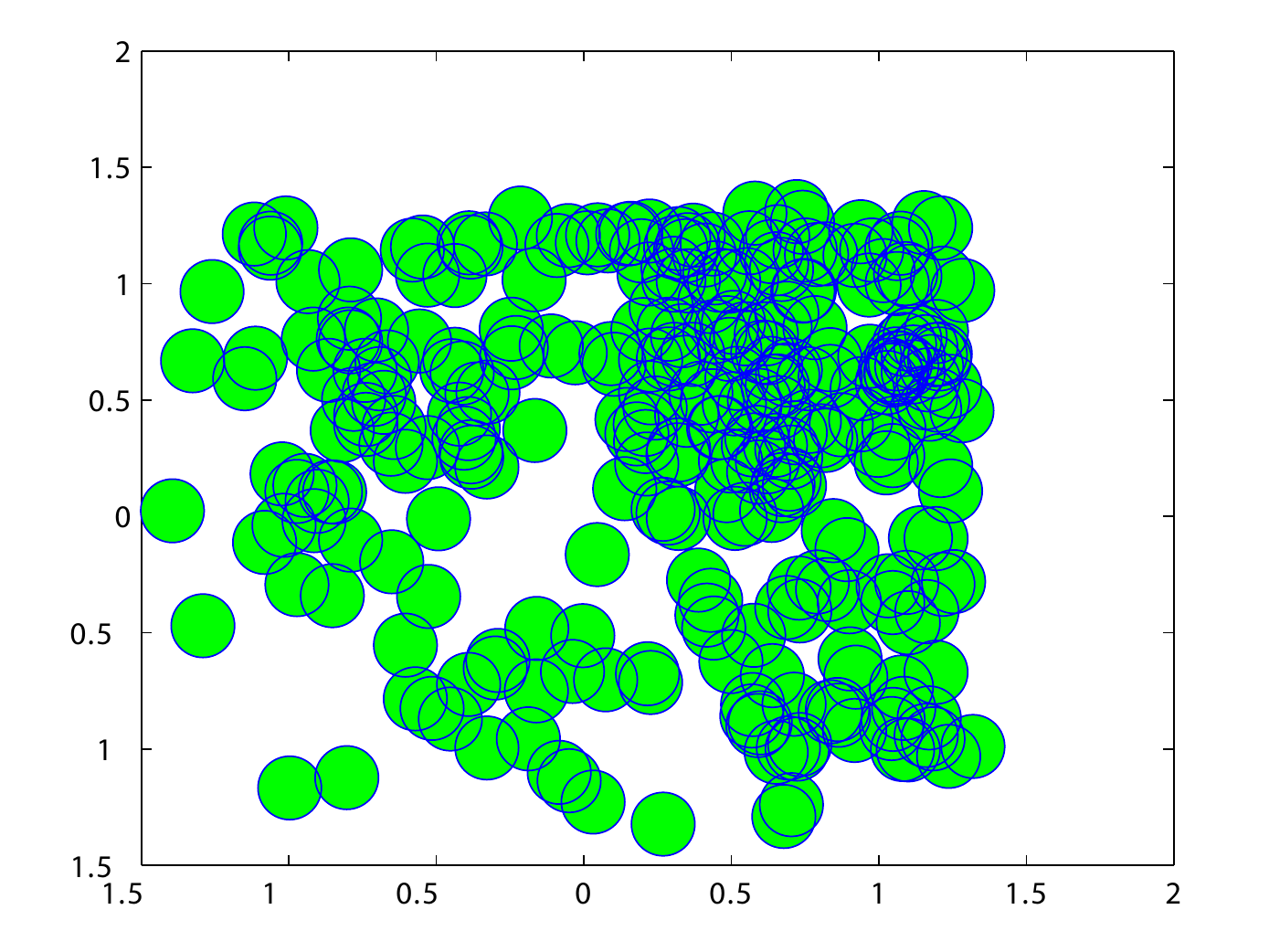}
\end{array}$
\caption{Two-dimensional projections of point clouds   corresponding to various instants of time; as time progresses, the value of the parameter $a$ increases (from top to bottom). Around each point, we draw a blue circle of a small radius (left column), and of a bigger radius (right column), to illustrate the birth and death of various loops. Color online.}
\label{fig:slices}
\end{figure}

First, we generate $d=4$ realizations of the $x$-time series  $\{x^k_n\}_n$, where $k=1,\ldots, d$, one for each value of the parameter $b=0.27$, $b=0.28$, $b=0.29$, $b=0.3$.  Thus, for each time $t_n$,
we have a point $x_n=(x^1_n,\ldots, x^d_n)\in\R^d$. For a sliding windows $w=50$, we generate a sequence of point cloud  data sets $X_n=(x_n, x_{n+1}, \ldots, x_{n+w-1})$. Hence    each cloud consists of $w$ points in  $\R^d$.
In Figure \ref{fig:slices}, we show two-dimensional projections of point clouds  corresponding to various instants of time; as time  progresses, the value of the parameter $a$ increases. Around each point, we draw a blue circle of a small radius (left column), and of a bigger radius (right column), to illustrate how loops are born and die. The panels at the top correspond to low values of the parameter $a$, when the deterministic H\'enon map   Eq. \eqref{eqn:henon} has a stable fixed point or a stable periodic orbit of low period; the corresponding noisy H\'enon map  Eq. \eqref{eqn:noisy_henon}  shows a cluster of points, or a few  clusters apart from one another.
The panels at the bottom correspond to larger values of the parameter $a$, when the deterministic H\'enon map Eq. \eqref{eqn:henon} undergoes a period doubling bifurcation that leads to a chaotic attractor; the corresponding noisy H\'enon map Eq. \eqref{eqn:noisy_henon}  shows a complex structure.

\begin{figure}
\centering\
\includegraphics[width=1\textwidth,height=0.45\textwidth]{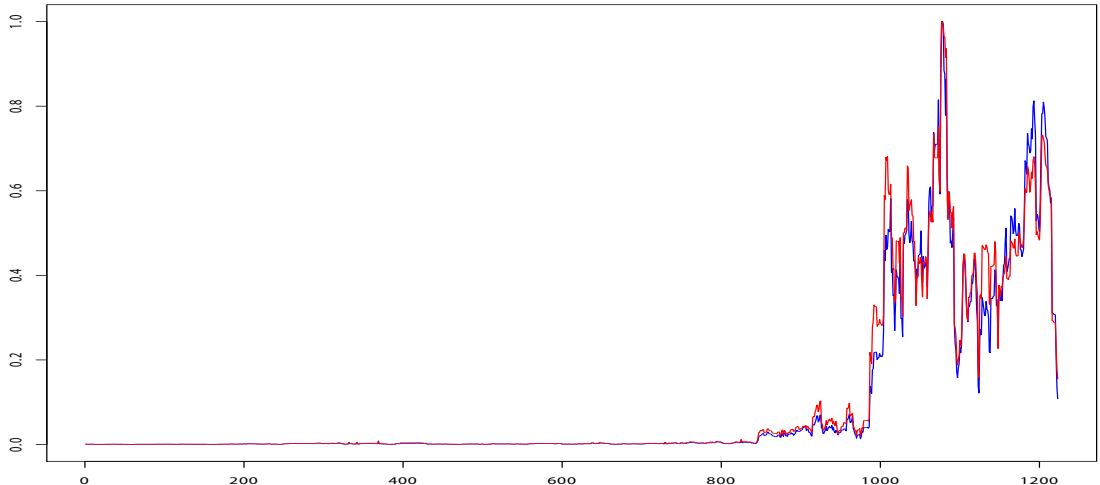}
\caption{Chaotic time series with noise. Plots of the normalized $L^1$-norm (blue) and $L^2$-norm (red) of the persistence landscapes; the horizontal axis corresponds to the
steps in time, and the vertical axis to the norms (see text for details). Color online.}
\label{fig:Henon_L1_L2_norm.pdf}
\end{figure}

Next, for each cloud point  $X_n$ we associate the corresponding Rips filtration $R(X_n,\eps)$, $\eps>0$, and we compute the $1D$  persistence diagram $P_1(X_n)$, the corresponding persistence landscape $\lambda(X_n)$, and the $L^1$ and $L^2$-norms: $\|\lambda(X_n)\|_1$ and $\|\lambda(X_n)\|_2$, respectively.
In Fig.~\ref{fig:Henon_L1_L2_norm.pdf} we show the normalized time series of
$\|\lambda(X_n)\|_1$ (in blue) and $\|\lambda(X_n)\|_2$ (in red). Both graphs show sharp increase around the parameter value $a$ that marks the onset of chaotic behavior.
We also point out that, when the $x$-time series move towards the chaotic regime, the  cross-correlation among them decays; in the chaotic regime, where there is sensitive dependence on initial conditions, the cross-correlation  typically gets below $\pm 10\%$. At the same time, the noise dressing added to the signal only plays a minor role.

The conclusion of this numerical experiment is that the  $L^p$-norms of persistence landscapes have the ability to detect transitions from regular to chaotic regimes, in systems with slowly evolving parameters. Passing from regular to chaotic dynamics determines significant changes in the topology of the attractor, which are picked up very clearly by the time series of  $L^p$-norms.
We stress that we only use this example to test out the proposed method. In the next subsection we consider a model that is more relevant to empirical financial time series.

\subsection{White noises with a growing variance}
The classic assumption made in theoretical quantitative finance is that equity prices follow a geometric Brownian motion with constant drift and diffusion coefficients. This model leads to normally distributed stock returns, defined as the difference in the logarithm of prices separated by a given time lag. In the following we use our method and perform Monte Carlo simulations of the  $L^p$-norms of persistence landscapes of a finite number $d=4$  white noises. Specifically, we wish to test whether a growing variance of the underlining white noises leads to increase in values of  $L^p$-norms.

\begin{figure}
\centering
\includegraphics[width=0.85\textwidth, height=0.45\textwidth]{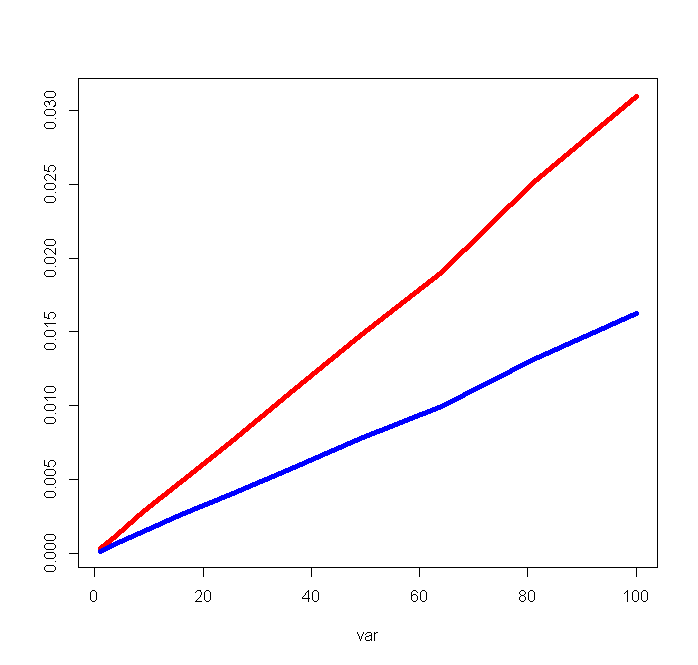}
\caption{Monte Carlo simulation, 100 realizations, of the dependency of the  $L^1$ (red line) and $L^2$ (blue line) norms of persistence landscapes on the growing variance of white noises, marked on the horizontal axis (see text for details). Color online.}
\label{fig:white_noise.png}
\end{figure}

We design our numerical experiment as follows. First, we generate four data sets, each with $100$ data points selected from Normal distributions $N(0,(\sigma+\delta_i)^2)$ with standard deviations $\sigma+\delta_i$, where $\sigma=1$   and $\delta_i$, $i = 1,\ldots,4$, has been chosen from the uniform distribution  $[-0.1, 0.1]$. Thereby, we construct a point cloud in the $4D$ space. We repeat this procedure multiple times. For each realization of the generated set $X_n$  we associate the corresponding Rips filtration $R(X_n,\eps)$, $\eps>0$, and compute the persistence diagram  $P_1(X_n)$, the corresponding persistent landscape  $\lambda(X_n)$, and the  $L^p$-norms: $\|\lambda(X_n)\|_1$  and  $\|\lambda(X_n)\|_2$.  We collect the values of the $L^p$-norms obtained at each realization, and calculate their mean-values at the end of the first simulation. We run this simulation $10$ times, with $\sigma$ that has been sequentially increased from $1$ to $10$, and find that the average values of the $L^1$-and $L^2$-norms of persistence landscapes rapidly converge with the number of iterations towards \emph{linearly increasing functions} of the average variance of white noises, see Fig. \ref{fig:white_noise.png}. This result is not surprising. When the average standard deviation   of white noises is increased by some scaling factor $F$, the effect on the persistence diagram is that each point $(b, d)$ becomes $(F\cdot b,F\cdot d)$, and the area  under the graph of each function $\lambda_k$ (see \eqref{eqn:landscape_2}) is changed by a factor of $F^2$. Hence, in this case, the values of the $L^p$-norms are proportional to the variance $\sigma^2$.

\subsection{White noises with Gamma-distributed inverse variance}

\begin{figure}
\centering
\includegraphics[width=1\textwidth,height=0.45\textwidth]{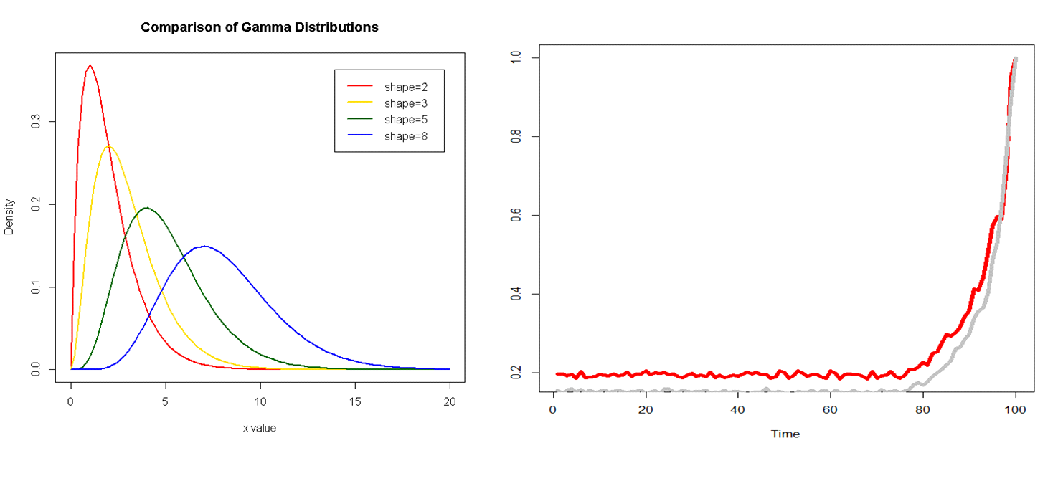}
\caption{(a)  Gamma-distributions with different values of the shape parameter; (b) Monte Carlo simulation of the time series of the normalized $L^1$-norm  (red line) of persistence landscapes derived for the process with the Gamma distributed inverse variance of the white noise. The gray line represents the time series of the normalized average variance of four underlining noises.  For details see text. Color online. }
\label{fig:gamma}
\end{figure}

Since early 1960's it has been known that PDFs of equity returns have ``fat tails" \cite{Mandelbrot,Fama}. The non-Gaussian distributions of relative changes in value of different asset types and on multiple time scales is an established stylized fact, see, e.g., \cite{Mantegna,Bouchaud} and the references therein. There is strong evidence that the time evolution of the volatility of stock returns is determined by its own stochastic process \cite{Shephard}.
There is a growing interest in modeling financial time series within the framework of non-extensive statistics \cite{Borland,Ausloos,Tsallis2003,Borland-Bouchaud,Biro,Tsallis2009} and `superstatistics' \cite{Gerig,Straeten,Vamos,Takaishi,Camargo,Katz2013a,Katz2013b}. The superstatistical approach \cite{Beck} significantly simplifies the description of the temporal evolution of the logarithm of asset returns, $x$.
Briefly, the model is based on the assumption that the typical relaxation time  $\tau$  of the local dynamics is much shorter than the time scale  $T_\gamma$ of externally driven fluctuations of its realized variance,  $\sigma^2=1/\gamma$. One assumes that a given discrete time series of $x$ containing $n$ data points can be divided into $N=n/T_\gamma$  time slices with locally equilibrium Normal distribution of $x$ and a distinct value of the realized local variance. Consequently, for  $N\gg 1$, the long-term, $t\gg T_\gamma$, the unconditional probability distribution of $x$ is approximated by the superposition of two distributions (therefore, superstatistics): locally normal distribution of $x$ and the probability distribution $f(\gamma)$  of the inverse variance
\begin{equation}\label{eqn:3.3}
p(x)=\int_0^\infty\sqrt{\gamma/2\pi}\exp(-\gamma x^2/2)f(\gamma)d\gamma.
\end{equation}
Marginalization over $\gamma$  expressed by this formula has been originally proposed by Praetz \cite{Praetz} who has shown that an infinite mixture of Gaussian PDFs of stock log-returns with the Gamma distributed precision  $\gamma$:
\begin{equation}\label{eqn:3.4}
 f(\gamma; \alpha, \beta)=\frac{\beta^\alpha}{\Gamma(\alpha)} \gamma^{\alpha-1}\exp(-\beta\gamma),\quad \gamma\geq 0,\alpha,\beta>0,
\end{equation}
yields the scaled Student's $t$-distribution, which provides a much better fit to empirical data than the normal distribution. Here  $\Gamma(\alpha)$ is the Gamma function,  $\alpha$ and $\beta$   are the shape and the rate parameters of the Gamma distribution, respectively. For large $\alpha$  the Gamma distribution converges to the Gaussian distribution with mean $\mu=\alpha/\beta$ and variance  $\sigma^2=\alpha/\beta^2$. In this limit, Eq.~\eqref{eqn:3.3} recovers the Normal distribution of log-returns, which follows from the classic geometric Brownian motion model of equity prices. Yet, for small values of the shape parameter corresponding to a high weight of realizations with high variance, see Fig.\ref{fig:gamma}-(a), Eq. \eqref{eqn:3.3} leads to fat-tailed distributions of $x$.

As Praetz  pointed out, by analogy with Brownian motion, the variance of price changes over a unit time interval (the diffusion coefficient) is proportional to a  slowly changing ``$\ldots$`temperature' of the share market $\ldots$, which represents the degree of activity or energy of the markets''. His insight and formulation of the problem are still incredibly valuable.
Recent studies of the daily time series of DJIA and S\&P 500 indices show that empirical distributions of $\gamma$  are close to the Gamma distribution \cite{Straeten,Micciche}.
Since the main objective of this paper is to explore the utility of TDA in forecasting of an approaching financial crash, which is conventionally associated with a growing `market temperature', in the following we use our method and perform Monte Carlo simulations of the  $L^p$-norms of persistence landscapes of $d=4$ white noises with Gamma-distributed inverse variance.

We generate $100$ point clouds in the $4D$ space, each with 100 data points selected from four Normal distributions $N(0,\gamma^{-1/2})$, with $\gamma$ that has been randomly taken from the set of $100$ Gamma-distributed variables.
To imitate the transition from the `cold' (low variance) to the `hot' (high variance) state of a market, for the first $75$ point clouds we keep the scale parameter relatively high  $\alpha=8$. Then we decrease this parameter by $0.25$ for each of the last $25$ steps. We keep the rate parameter fixed,  $\beta=1$. For each generated point cloud  $X_n$ we associate the corresponding Rips filtration $R(X_n,\eps)$, $\eps>0$, and compute the $1D$ persistence diagram  $P_1(X_n)$, the corresponding persistence landscape $\lambda(X_n)$, and the norms:  $\|\lambda(X_n)\|_1$ and $\|\lambda(X_n)\|_2$, respectively. We repeat this procedure multiple times, collect the values of $L^p$-norms obtained at each realization, and calculate the relevant mean-values.   In Fig. \ref{fig:gamma}-(b) we show the normalized time series of the $L^1$-norm and the average variance of underlining noises, averaged over $100$ realizations; the behavior of the $L^2$-norm is almost indistinguishable from that of the $L^1$-norm and is not shown here.
The plot shows a sharp increase of the  $L^1$-norm of persistence landscapes with the decrease of the shape parameter, which corresponds to a growing variance of the distributions of $x$. This numerical experiment demonstrates that  the $L^p$-norms of persistence landscapes are very sensitive to transitions in the state of a system from a regular to a `heated' regime.
Note that the cross-correlation between the generated noises at each realization is rather low: it never exceeds $\pm 20\%$. On the other hand, we do not observe any persistent loops for identical noises. In this special case, the $L^p$-norms are always equal to zero.

Once again, we stress that the goal of these simulations is to check the applicability of the proposed method to noisy multidimensional time series that are typical for complex systems. We certainly do not claim that the particular models considered here are representative for how crises occur in financial markets.

\section{Empirical analysis of financial data}
\label{sec:empirical}

%Interpret loops in terms of stock returns

\begin{figure}
\centering\
\includegraphics[width=1.0\textwidth]{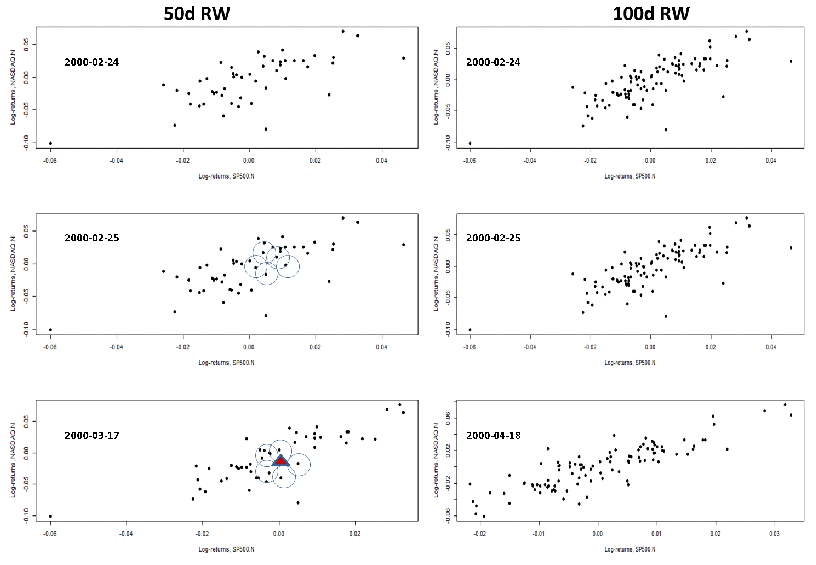}
\caption{Scattered plots of the time series of the normalized daily log-returns of S\&P 500 and NASDAQ within 50 and 100 days intervals ending at selected dates. The left column schematically demonstrates birth of the $1D$ homology (loops) at
certain scale (radius) in the  $2D$ point cloud of $50$ data points.
}
\label{fig:scattered}
\end{figure}

We analyze the daily time series of four major US stock market indices: S\&P 500, DJIA, NASDAQ, and Russel 2000 between December 23, 1987 and December 08, 2016 ($7301$ trading days). The times series data were downloaded from Yahoo Finance. For each index and for each trading day we calculate log-returns defined as the forward daily changes in the logarithm of the ratio $r_{ij}=\ln(P_{i,j}/P_{i-1,j})$, where $P_{i,j}$  represents the adjusted closing value of the index $j$ at the day $i$. We wish to explore the topological properties of this multi-dimensional time series. In our approach, the point cloud is formed by $w$  points in $\R^d$, where the coordinates of each point in $\R^d$ represent the daily log-returns. That is, each point cloud is given by a $w\times d$ matrix, where the number of columns $d = 4$ is the number of $1D$ time series involved in our analysis, and the number of rows $w$ is the size of the sliding window. We consider two sizes of the window: $w=50$ and $w=100$ trading days; see Fig. \ref{fig:scattered} for an example of $2D$ point clouds. The sliding step is set to one day, which in the case under consideration yields a $(7300 - w)$ time-ordered set of point clouds.

\begin{figure}
\centering$\begin{array}{c}
\includegraphics[width=0.84\textwidth]{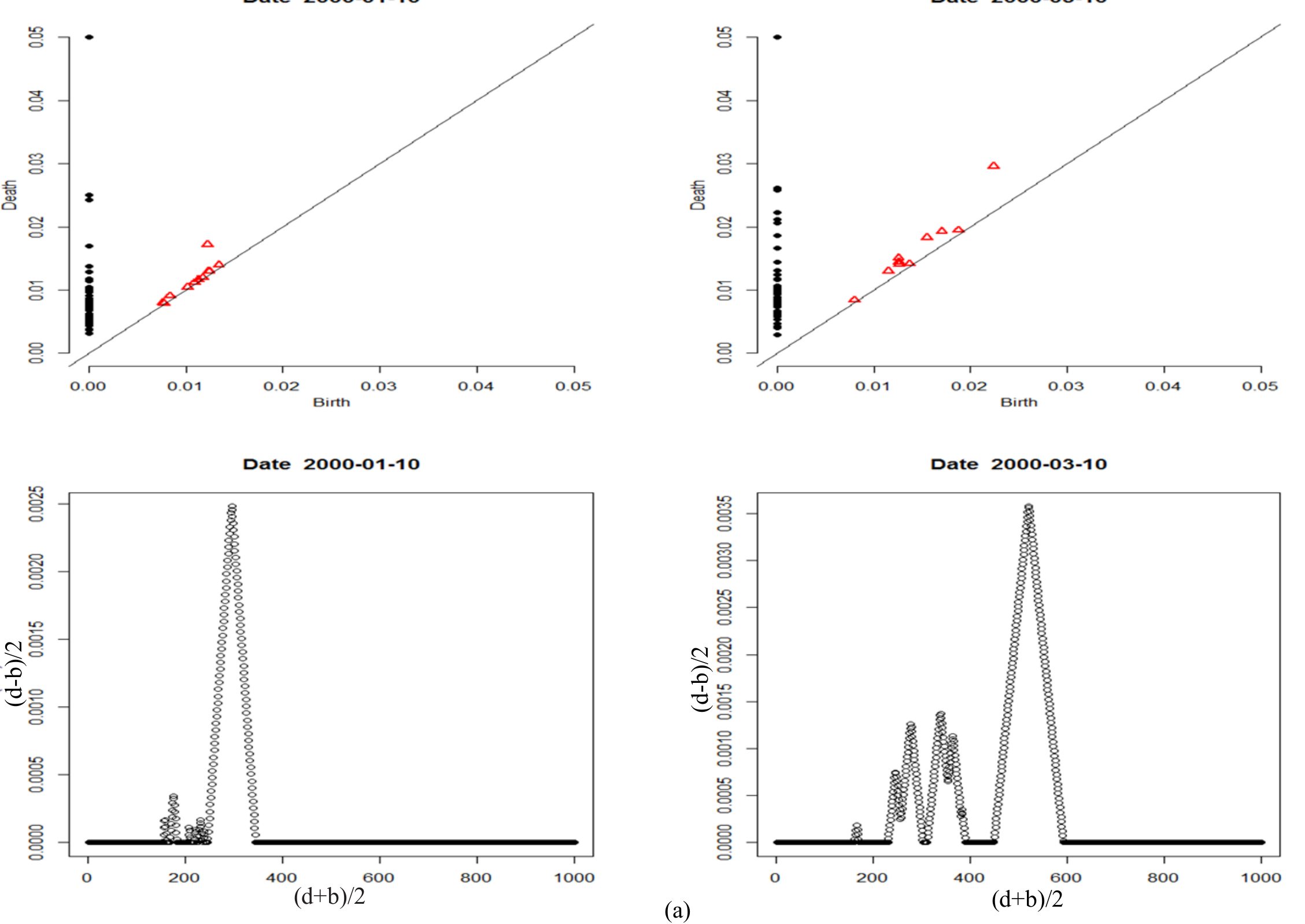}\\
\includegraphics[width=0.84\textwidth]{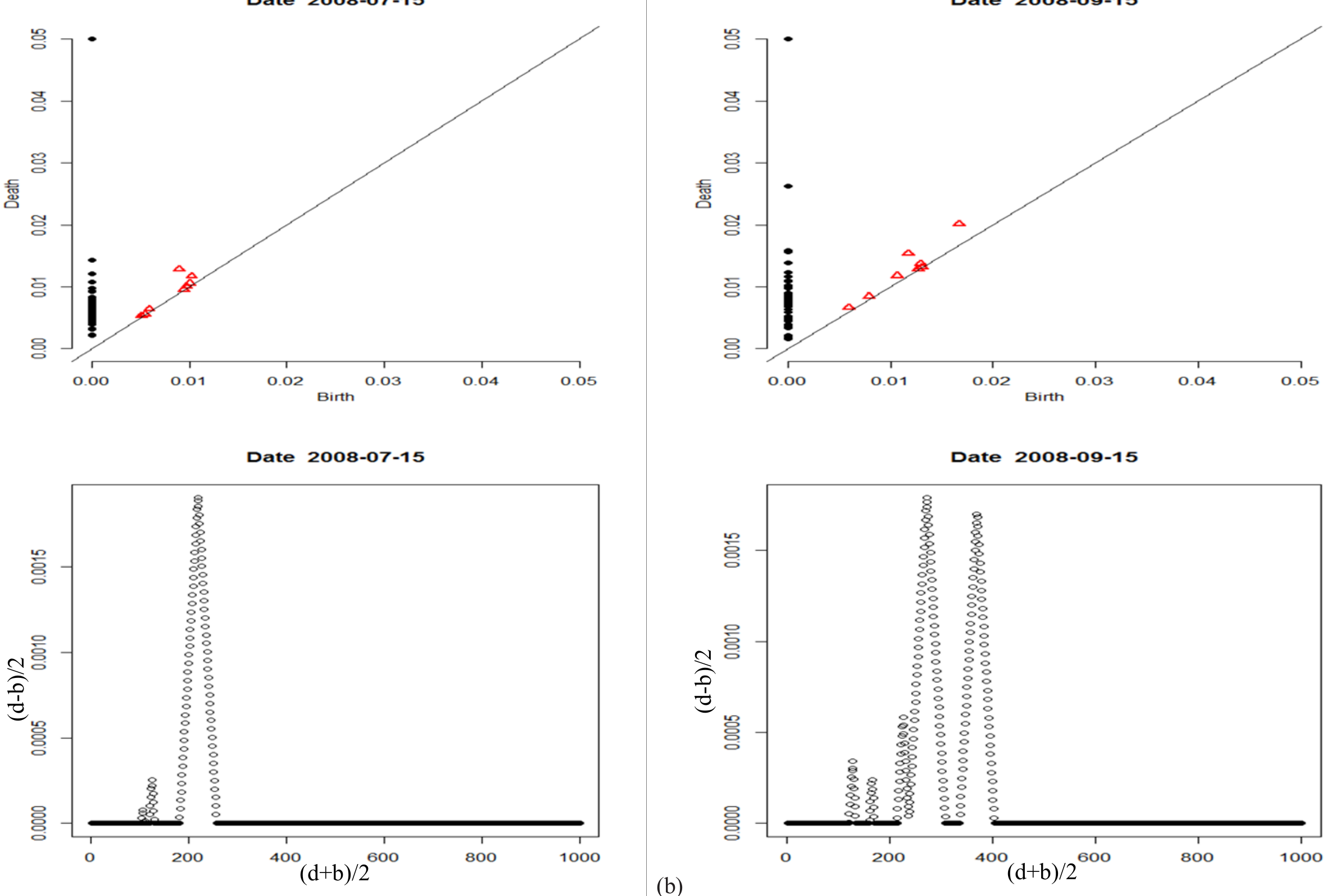}
\end{array}$
\caption{The Rips persistence diagrams and the corresponding persistence landscapes calculated with the sliding window of $50$ trading days on selected dates. The solid black dots represent connected components, red triangles represent loops. (a) Technology crash, 2000; (b) Financial Crisis, 2008. Color online. }
\label{fig:Rips}
\end{figure}

Figure \ref{fig:Rips} shows the Rips persistence diagrams and the corresponding $1$-st landscape function $\lambda_1$, defined by Eq. \eqref{eqn:landscape_2}, calculated with the sliding window of 50 trading days at certain dates. The $x$ and $y$ coordinate of Rips persistence diagrams correspond to the times of birth ($b$) and death ($d$) of a topological feature, respectively. The $x$ and $y$ coordinate of persistence landscapes correspond to the rescaled axes: $x=(d+b)/2$, and $y=(d-b)/2$. In plain words, the x-coordinate of a persistence landscape is the average parameter value over which the feature exists, and the y-coordinate is the half-life of the feature.
Figure \ref{fig:Rips} demonstrates a visually clear separation of topological signals from a topological noise per point cloud. Two of selected dates correspond to the date of the technology crash on March 10, 2000 (\ref{fig:Rips}-(a), right column) and the date of the Lehman bankruptcy on 09/15/2008 (Fig. \ref{fig:Rips}-(b), right column).
It is clearly seen from these plots that as the stock market becomes more volatile loops in the relevant point clouds become much more pronounced.

\begin{figure}
\centering$\begin{array}{c}
\includegraphics[width=1\textwidth,height=0.45\textwidth]{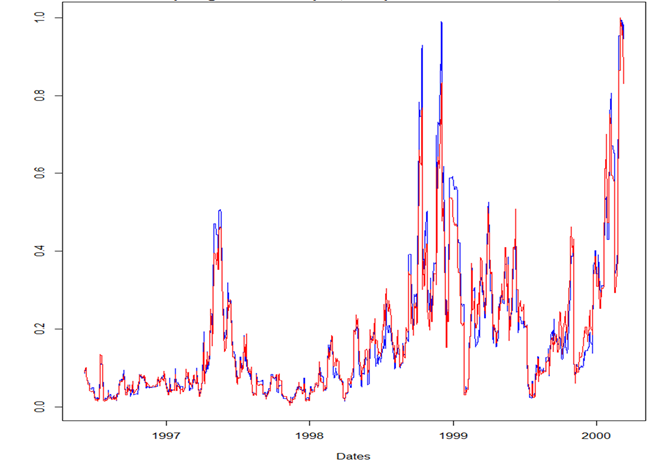}
\end{array}$
\caption{The time series of normalized $L^1$ (blue line) and $L^2$ (red line) norms of persistence landscapes calculated with the sliding window of 50 days.  Color online.}
\label{fig:Crashes-c}
\end{figure}

To quantify this behavior, as a convenient summary of changes in topological features throughout filtration, we calculate the $L^p$  norms ($p=1,2$) of the persistence landscapes $\lambda$  for each rolling window. The set of these values forms the daily time series of these measures; see Fig. \ref{fig:Crashes-c} for an example.

\begin{figure}
\includegraphics[width=1\textwidth]{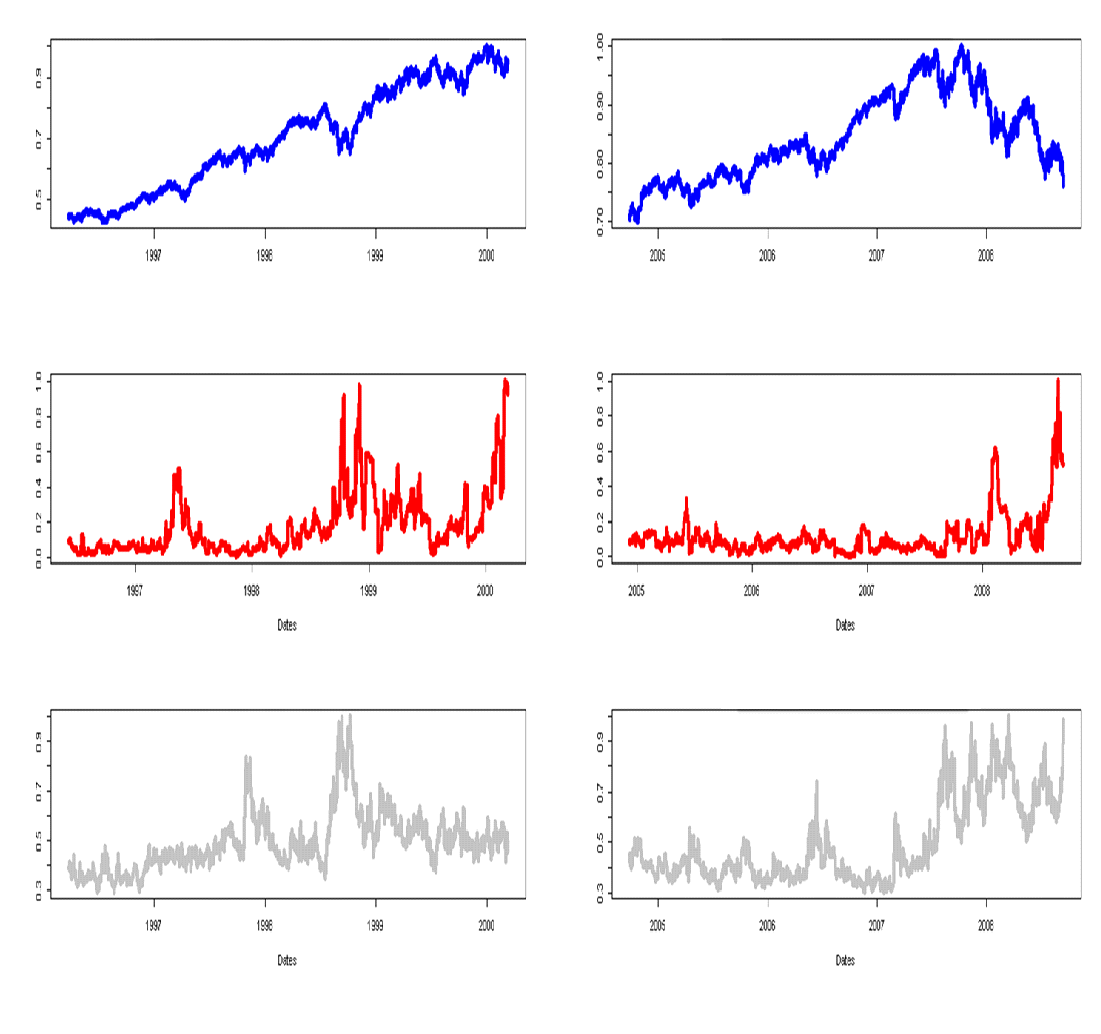}
\caption{$1000$ trading-days prior to a crash. Top-down, the daily time series of the normalized S\&P 500, the normalized  $L^1$-norm of topological landscapes calculated with the sliding window of $50$ days, and VIX. Left column: proximity of the `dotcom crash' on 03/10/2000.  Right column: proximity of the Lehman bankruptcy on 09/15/2008.}\label{fig:1000days}

\end{figure}

Figure \ref{fig:1000days} shows the normalized time series of the $L^1$ norms of persistence landscapes calculated with the sliding window of $w=50$ trading days around the date of the technology crash and the date of the Lehman bankruptcy.
For comparison, we also show the corresponding segments of the normalized time series of S\&P 500 and the volatility index VIX. The VIX measures the implied volatility in the S\&P 500 index over the next $30$ days and is often used for forecasting purposes. It is qualitatively clear from these plots that the time series of the  $L^1$-norm demonstrates strong fore-shocks prior to the primary peak, which mounts during a crisis. The change of the size of the rolling window from $50$ to $100$ days as well as the temporal behavior of the  $L^2$-norm   closely resemble this picture (not shown here). The only notable differences in the outcome related to the change of the window size are an overall flattening of spikes in the time series of the  $L^p$-norms and a slight shift of the primary maximum of these time series towards the later dates.

\begin{figure}
\centering$\begin{array}{c}
\includegraphics[width=0.75\textwidth]{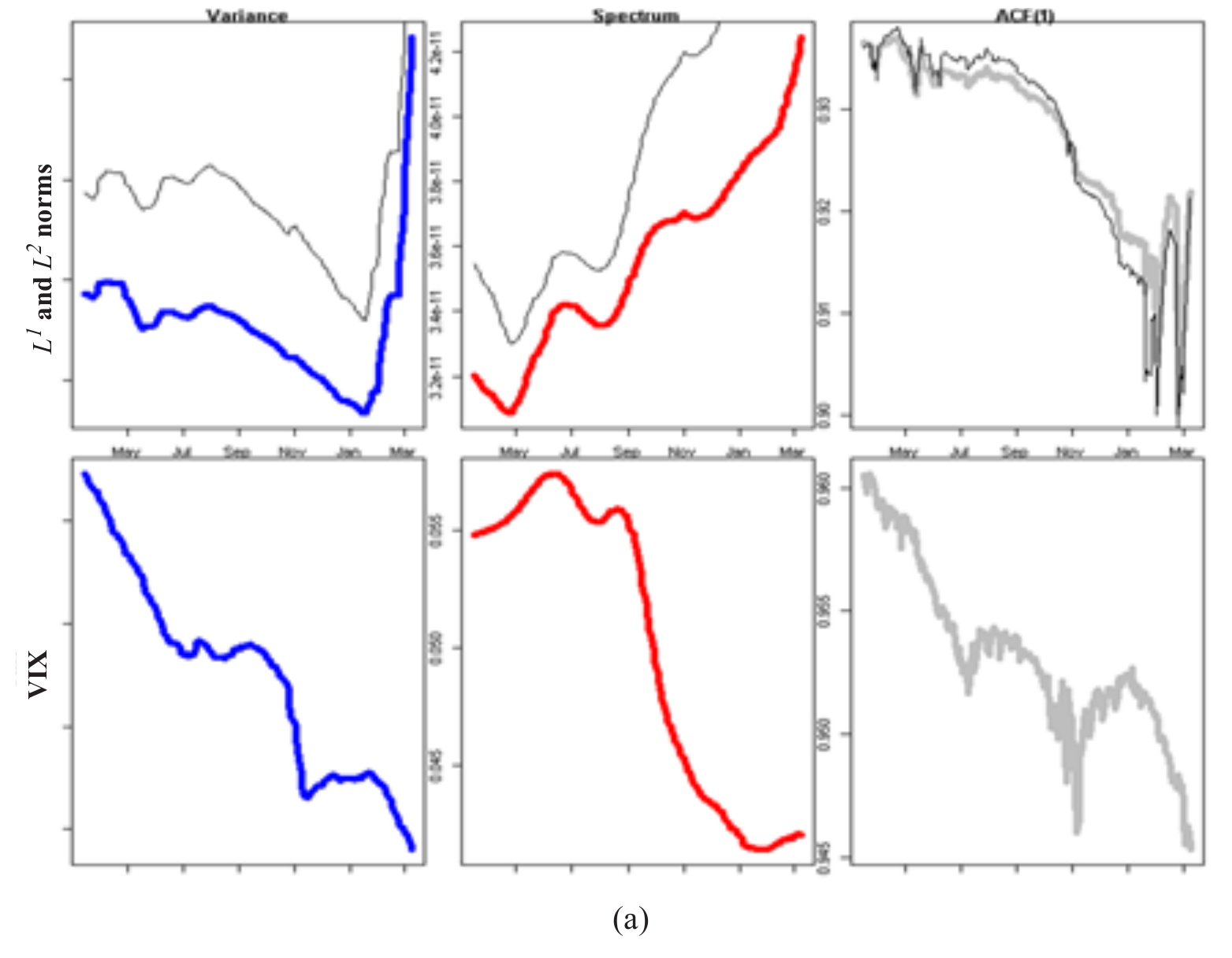}\\
\includegraphics[width=0.72\textwidth]{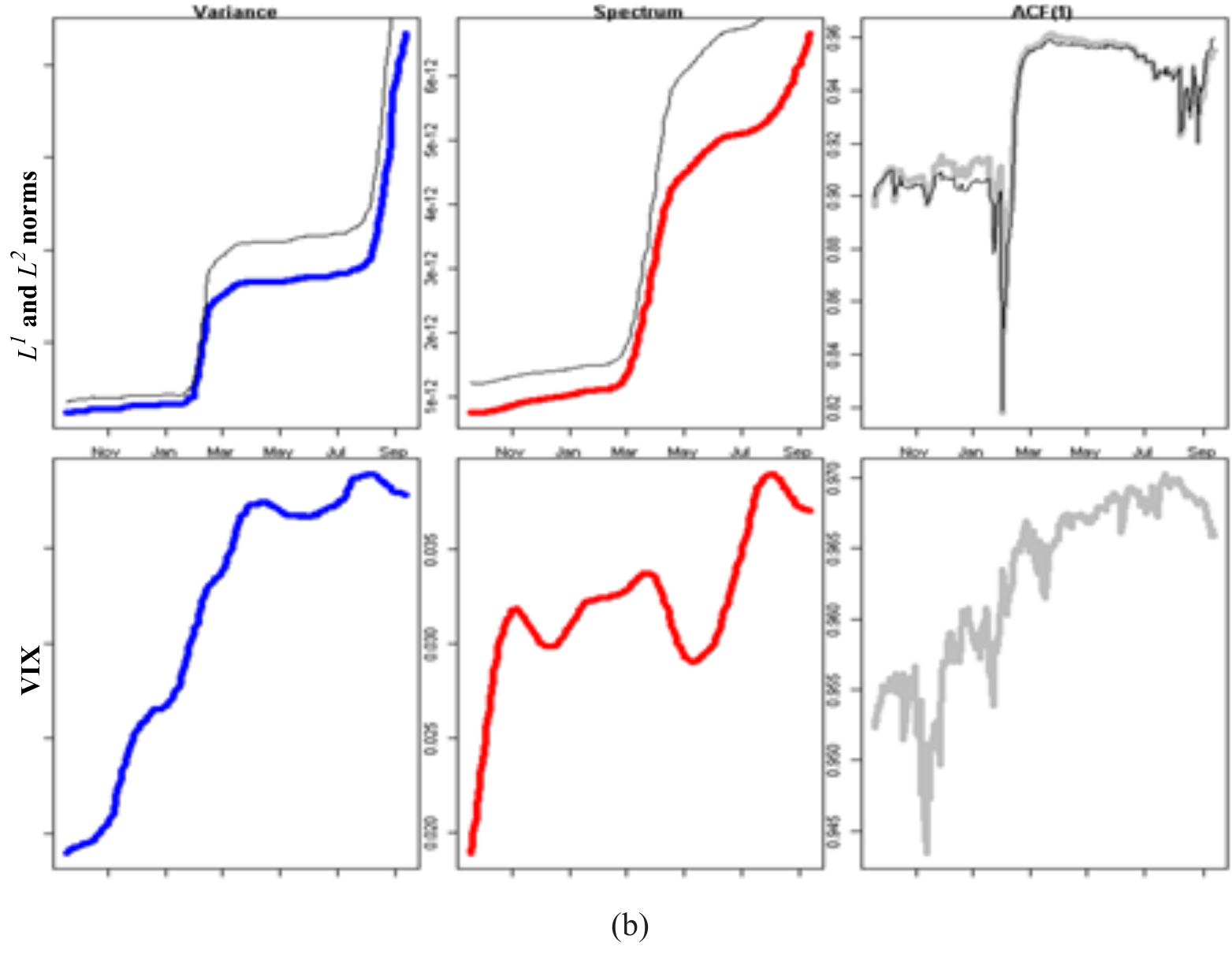}
\end{array}$
\caption{The variance, the average spectral density at low frequencies, and the first lag of the ACF of the time series of $L^p$ norms of persistence landscapes, and VIX for 250 trading-days prior to (a)~the technology crash on 03/10/2000; (b)~the Lehman bankruptcy on 09/15/2008 (see text for details). The thin black line corresponds to $L^1$.}
\label{fig:Kendall}
\end{figure}

	Having the time series of $L^p$ norms of persistence landscapes in place, we apply the standard technique and analyze the leading statistical indicators of these time series. In that way, we further quantify temporal variations in the persistence  of loops in the time-ordered set of point clouds with a growing market stress. We follow Guttal et.al. \cite{Guttal16}, and employ the $500$ days rolling window with the sliding step of one day to calculate the variance, the average spectral density at low frequencies, and the first lag of the auto-correlation function (ACF) for $250$ trading days prior to the date of a market crash. This window should not be confused with the sliding window of $w=50$ days that has been used for encapsulation of point clouds throughout the computation of the time series of $L^p$-norms of persistence landscapes. Figure \ref{fig:Kendall} shows the obtained results. It is clearly seen from the plots on this figure that the variance and the average spectral density at low frequencies of the time series of $L^p$-norms are substantially growing prior to any of the crashes. We quantify the observed trend with the Mann-Kendall test, which statistically assess a monotonic upward or downward movement of the variable over time \cite{Kendall75,McLeod11}.  The strong trend of rising average spectral density at low frequencies for $250$ trading days prior to either dotcom crash on 03/10/2000 or to the Lehman Bankruptcy on 09/15/2008 is especially pronounced: Kendall-tau rank correlation coefficients are $0.89$ and $1.00$, respectively. On the other hand, the ACF at lag-$1$ shows no trend prior to any of these events. We note that doubling the size of the point cloud to $100$ data points does not significantly change this picture (not shown here). For comparison, we also present on Fig. \ref{fig:Kendall} the temporal behavior of the same statistical indicators obtained with the same rolling window of $500$ days from the daily time series of the volatility index. Interestingly, prior to the 2000 crash the time series of VIX shows a declining variability. On the other hand, similarly to the norms of persistence landscapes, the average spectral density at low frequencies of the time series of VIX demonstrates strong rising trend for $250$ trading days prior to the Lehman Bankruptcy (Kendall-tau $= 0.89$).

\section{Conclusions}\label{sec:conclusions}
We apply TDA for exploration of the temporal behavior of topological features in the well-known financial data as the state of the market is changing with time. The novelty of our approach to multi-source time series is twofold: (i) we use the single sliding window to enclose the point cloud in the $d$-dimensional space determined by the number of time series under consideration; (ii) we employ the  $L^p$-norm of persistence landscapes as a quantifier of stability of topological features. We believe that this approach can be used beyond the analysis of financial time series presented here.

More specifically, we study persistence of loops ($1D$ persistence homology) in a $4D$ point cloud formed by a \emph{single} sliding window and \emph{four} $1D$ time series of daily log-returns of the four stock-market indices. Thus, contrary to the traditional two-parametric time-delay embedding method, our approach has only one parameter: the size of the window, $w$. Our study reveals that the \emph{shape} of financial time series strongly depends on the state of the market. The empirical analysis shows that the time series of the $L^p$-norms exhibit strong growth around the primary peak emerging during a crisis. This behavior reflects an increased persistence of loops appearing in point clouds as the market undergoes transition from the ordinary to the ``heated" state. Remarkably, the variability of the time series of the  $L^p$-norms of persistence landscapes, quantified by the variance and the spectral density at low frequencies, demonstrates a strong rising trend for $250$ trading days prior to either dotcom crash on 03/10/2000 or to the Lehman Bankruptcy on 09/15/2008. Our study suggests that TDA offers a  novel econometric method, and is yielding the new category of early warning signals of an imminent market crash.

\section{Acknowledgement}
Research of M.G. was partially supported by NSF grant  DMS-0635607 and by the  Alfred P. Sloan Foundation grant G-2016-7320. Research of Y.K. was supported by S\&P Global Market Intelligence. The first author is grateful  to  K.~Mischaikow for useful discussions.
The views expressed in this paper are those of the authors, and do not necessary represent the views of S\&P Global Market Intelligence.

\end{document}